\documentclass{aa}
\usepackage{graphicx}      
\usepackage{txfonts}
\usepackage{natbib}
\usepackage{epsfig}
\usepackage{subfigure}
\bibpunct{(}{)}{;}{a}{}{,} 

\begin{document}
   \title{Molecular gas heating in Arp 299\thanks{{\it Herschel} is an ESA space observatory with science instruments provided by European-led Principal Investigator consortia and with important participation from NASA.}
   \fnmsep \thanks{The \emph{Herschel}/SPIRE spectra are available in electronic form
at the CDS via anonymous ftp to cdsarc.u-strasbg.fr (130.79.128.5)
or via http://cdsweb.u-strasbg.fr/cgi-bin/qcat?J/A+A/}}

   \author{M.~J.~F.~Rosenberg
          \inst{1}
          \and
          R. Meijerink \inst{1,2}
          \and
          F.~P. Israel \inst{1}
          \and
          P.~P. van der Werf \inst{1}
          \and
          E.~M.~Xilouris \inst{3}
          \and
          A. {Wei{\ss}} \inst{4}
          }

          \institute{Leiden Observatory, Leiden University,
             P.O. Box 9513, 2300 RA Leiden, The Netherlands\\
            \email{rosenberg@strw.leidenuniv.nl} \and University of
            Groningen, Kapteyn Astronomical Institute, University of
            Groningen, PO Box 800, 9700 AV Groningen, The Netherlands
            \and Institute for Astronomy, Astrophysics, Space Applications \& Remote 
            Sensing, National Observatory of Athens, P. Penteli, 15236 Athens, Greece
            \and Max-Planck-Institut für Radioastronomie,\\
             Auf dem Hügel 16, Bonn, D-53121, Germany}

   \date{}

   \abstract 
   {(Ultra) luminous infrared galaxies ((U)LIRGs) are nearby
     laboratories that allow us to study similar processes to those
     occurring in high redshift submillimeter galaxies.  Understanding
     the heating and cooling mechanisms in these galaxies can give us
     insight into the driving mechanisms in their more distant
     counterparts.  Molecular emission lines play a crucial role in
     cooling excited gas, and recently, with Herschel Space
     Observatory we have been able to observe the rich molecular spectrum.
     Carbon monoxide (CO) is the most abundant and one of the brightest
     molecules in the Herschel wavelength range.  CO transitions from
     J=4-3 to 13-12 are observed with Herschel, and together, these
     lines trace the excitation of CO. We study Arp 299, a
     colliding galaxy group, with one component (A) harboring an AGN and two more (B and C) 
     undergoing intense star formation.
     For Arp 299 A, we present PACS spectrometer observations of high-J CO lines up to J=20-19 and JCMT observations of $^{13}$CO and HCN to discern between
     UV heating and alternative heating mechanisms. There is an immediately noticeable difference
     in the spectra of Arp 299 A and Arp 299 B+C, with source A having 
     brighter high-J CO transitions. This is reflected in their respective spectral energy line distributions. We find that photon-dominated regions (PDRs, UV heating) are unlikely to heat all the gas since 
     a very extreme PDR is necessary to fit the high-J CO lines.  In addition, this extreme PDR does not fit the HCN observations, and the dust 
     spectral energy distribution shows that there is not enough hot dust to match the amount expected from such an extreme PDR.   Therefore, we determine that the high-J CO and HCN transitions are heated
     by an additional mechanism, namely cosmic ray heating, mechanical heating, or X-ray heating.  We find that mechanical heating, in combination with UV heating, is the only mechanism that fits all molecular transitions.
     We also constrain the molecular gas mass of Arp 299 A to $3\times10^9$ M$_\odot$ and find that we need
     4\% of the total heating to be mechanical heating, with the rest UV heating.  Finally, we 
     caution against the use of $^{12}$CO alone as a probe of physical properties in the interstellar medium. }

\titlerunning{CO excitation in Arp 299}
\authorrunning{Rosenberg, M.~J.~F. et al.}

   \maketitle
%

\section{Introduction}
(Ultra) luminous infrared galaxies ((U)LIRGs) are systems or galaxies
with very high far-infrared luminosity (ULIRG: L$_{FIR} >
10^{12}L_\odot$ and LIRG: L$_{FIR} > 10^{11}L_\odot$;
\cite{1996ARA&A..34..749S}) owing to a period of intense star formation.
Arp 299 (NGC 3690 + IC 694, Mrk 171, VV 118, IRAS 11257+5850, UGC6471/2) is a
nearby (42 Mpc \cite{1991ApJ...366L...1S}) LIRG (L$_{FIR} =
5\times10^{11}$) currently undergoing a major merger event. Arp 299 is dominated by
intense, merger-induced star formation and is made up of three main
components \citep{2000ApJ...532..845A}.  Although the core regions of these
components can still be resolved, there is a large overlap in their disks.
The separation between Arp 299 A and Arp 299 B and C is 22'', or 4.5 kpc in physical
distance.  Arp 299 B and C are separated by only 6.4'', or 1.4 kpc.  The largest component is the massive galaxy IC 694 (Arp 299 A),
which accounts for about 50\% of the galaxies' total infrared luminosity
\citep{2000ApJ...532..845A}.  

The galaxy NGC 3690 represents
the second component (Arp 299 B) that is merging into IC 694 and
represents $\sim 27\%$ of the total luminosity
\citep{2000ApJ...532..845A}.  The third component (Arp 299 C) is an
extended region of star formation where the two galaxy disks overlap. Here we use the standard nomenclature, instead of 
the NED definition.
\citet{1991ApJ...366L...1S} suggest that an active galactic nucleus
(AGN) could be responsible for the large amount of far-infrared
luminosity in Arp 299 A, although \citet{2000ApJ...532..845A} find no
supporting evidence.  \citet{2005A&A...436...75H} and \citet{2007NewAR..51...67T}
suggest that the presence of H$_2$O masers, along with X-ray
imaging and spectroscopy
\citep{2002ApJ...581L...9D,2003ApJ...594L..31Z,2004ApJ...600..634B}
indicate that an AGN must be present in the nuclear region of Arp 299 A.
Using milliarcsecond 5.0 GHz resolution images from the VLBI, \citet{2010A&A...519L...5P}
conclude that there is a low luminosity AGN (LLAGN) at the center of
Arp 299 A.

In addition to the AGN, there are intense knots of star formation observes
in the infrared and radio \citep{1991ApJ...377..426W}. \citet{2000ApJ...532..845A} observes Arp 299 in 
high-resolution with the
Hubble Space Telescope in the near-infrared and also find that
over the past 15 Myr, Arp 299 has been undergoing intense merger-related star formation.
This star formation is fueled by large amounts of dense molecular gas: 8$\times10^5$ M$_\odot$ pc$^{-2}$ 
for Arp 299 A, 3$\times10^4$ M$_\odot$ pc$^{-2}$ for Arp 299 B, and 2$\times10^4$ M$_\odot$ pc$^{-2}$ 
for Arp 299 C \citep{1991ApJ...366L...1S}. Most of the star formation responsible for the high 
far-infrared luminosity is spread over 6-8 kpc \citep{2009ApJ...697..660A}, resulting in most of Arp 299 having
typical starburst properties.  Only the nucleus of Arp 299 A exhibits true LIRG conditions, 
with n$_e$=1-5$\times10^{3}$ cm$^{-3}$, deep silicate absorption features implying embedded 
star formation, and PAH emission \citep{2009ApJ...697..660A}.  

In this paper we present observations of the central region of Arp 299
using the Spectral and Photometric Imaging Receiver (SPIRE) on board
of the ESA \emph{Herschel} Space Observatory as part of HerCULES (PI:
P.~P. van der Werf).  Due to the large spectral range of SPIRE, we can
observe many different line transitions, which enables the study of
excitation mechanisms of different phases of the ISM.  Specifically,
we compare the intensity of different CO transitions to CO
emission models to determine the density, temperature, and radiation
environment of the phases of the ISM in Arp 299.  We directly
compare Arp 299 A, which harbors an AGN, to Arp 299 B and C, which are
undergoing rapid star formation. Then we add observations from the Photodetector Array Camera and Spectrometer (PACS) (PI: R. Meijerink) and the literature to disentangle the heating mechanisms
of the molecular gas.  In Section~\ref{sec:obs} we present all of the observations and discuss the data
reduction methods.  Then in Section~\ref{sec:res}, we present the spectra and line fluxes for both the SPIRE and 
PACS spectra. A qualitative comparison between Arp 299 A, B, and C is discussed in Section \ref{sec:comp}.  Using 
all available data, in Section~\ref{sec:case} we explore the heating mechanisms of the highest-J CO transitions and 
discuss the limitations of using only $^{12}$ CO to determine physical parameters in Section~\ref{sec:limit}.  We 
state our conclusions in Section~\ref{sec:conc}.

\section{Observations and data reduction}
\label{sec:obs}
\subsection{Observations}

\begin{table}
  \caption[]{Log of Herschel Observations}
\label{OBS_IDS}    
\begin{center}
\resizebox{8cm}{!}{
\begin{tabular}{l c c c c } 
\vspace{-5.0mm}    \\
\hline\hline   
\noalign{\smallskip}
Instru-   &  Transition & Observation ID  &  Date   & Integr. \\
ment      &             &        &  Y-M-D  & [s]         \\
\noalign{\smallskip}
\hline               
\noalign{\smallskip}
\multicolumn{5}{c}{Arp 299 A $11^h28^m33^s.7$ $+58^d33^m46^s$} \\
\noalign{\smallskip}
\hline               
\noalign{\smallskip}
PACS         & CO $J=14-13$     & 1342232607 & 2011-11-21 & 4759 \\
PACS         & CO $J=16-15$     & 1342232606 & 2011-11-21 & 641  \\
PACS         & CO $J=18-17$     & 1342232608 & 2011-11-22 & 782  \\
PACS         & CO $J=20-19$     & 1342232603 & 2011-11-21 & 1225 \\
PACS         & CO $J=22-21$     & 1342232605 & 2011-11-21 & 976  \\
PACS         & CO $J=24-23$     & 1342232603 & 2011-11-21 & 1225 \\
PACS         & CO $J=28-27$     & 1342232607 & 2011-11-21 & 4759 \\
SPIRE        & $194-671$~$\mu$m & 1342199248 & 2011-06-27 & 4964 \\
\noalign{\smallskip}
\hline               
\noalign{\smallskip}
\multicolumn{5}{c}{Arp 299 B $11^h28^m31^s$ $+58^d33^m41^s$} \\
\noalign{\smallskip}
\hline               
\noalign{\smallskip}
SPIRE        & $194-671$~$\mu$m & 1342199249 & 2011-06-27 & 4964 \\
\noalign{\smallskip}
\hline
\noalign{\smallskip}
\multicolumn{5}{c}{Arp 299 C $11^h28^m31^s.13$ $+58^d33^m48^s.2$} \\
\noalign{\smallskip}
\hline
\noalign{\smallskip}
SPIRE        & $194-671$~$\mu$m & 1342199250 & 2011-06-27 & 4964 \\
\noalign{\smallskip}
\hline               
\noalign{\smallskip}
\end{tabular}}
\end{center}
\end{table}

{\it Herschel SPIRE FTS data:} Observations of Arp 299 were taken with
the {\it Herschel} Spectral and Photometric Imaging Receiver and
Fourier-Transform Spectrometer \citep[SPIRE-FTS,][]{2010A&A...518L...3G} on board
the Herschel Space Observatory \citep{2010A&A...518L...1P} using three
separate pointings centered on Arp 299 A, Arp 299 B, and Arp 299 C (see
Table \ref{OBS_IDS}). The low frequency
band covers $\nu$=447-989 GHz ($\lambda$=671-303 $\mu$m) and the high
frequency band covers $\nu$=958-1545 GHz ($\lambda$=313-194 $\mu$m),
and these bands include the CO J=4-3 to CO J=13-12 lines. The high spectral resolution mode was used with
a resolution of 1.2 GHz over both observing bands. Each source was
observed for 4964 seconds (1.4 hours).  A reference measurement was
used to subtract the emission from the sky, telescope, and instrument.  We present the original observed SPIRE spectra in 
Figure~\ref{fig:spectra}.

{\it Herschel SPIRE Photometry data:} Observations using the SPIRE Photometer 
were taken as part of the Herschel Guaranteed Time Key Program SHINING (PI: E. Sturm).
The system was observed on the 6th of January 2010 at 250, 350, and 500 $\mu$m (observation ID: 1342199344, 1342199345, 1342199346). The source was observed 797 seconds in total.

{\it Herschel PACS spectroscopy data:} CO $J_{up} \geq 14$ observations were made
with the Photodetector Array Camera and Spectrometer
\citep[PACS,][]{2010A&A...518L...2P} for Arp 299 A only. The data presented
here have been obtained as part of the Herschel program OT1\_rmeijeri\_1
(PI: Meijerink), complemented by observations from OT1\_shaileyd\_1
(PI: Hailey-Dunsheath). The observations consisted of deep
integrations targeting CO $J = 14-13$, CO $J = 16-15$, CO $J =
18-17$, CO $J = 20-19$, CO $J = 22-21$, CO $J = 24-23$, and CO $J =
28-27$. The observation IDs of the targeted CO lines are listed in
Table \ref{OBS_IDS}. 

{\it Ground based data:} We use the short spacing
corrected CO maps from \citet{2012ApJ...753...46S} for the J=1-0, 2-1, and
3-2 transitions for Arp 299 B and C.  We integrate the flux corresponding to our largest
SPIRE beam (J=4-3, 42'') full-width-half-maximum (FWHM) centered on each of the pointings
respectively.  We do not use these values for Arp 299 A since the CO 1-0 map has 
error bars larger than 50\%. 

For Arp 299 A, we used dual-polarisation receivers A and B (decommissioned in 2009)
on the IRAM 30 m telescope to measure the $J$=1-0 $^{12}$CO line
towards Arp~299 in November 2005, followed by observations of $J=2-1$
$^{12}$CO and both $J$=1-0 and $J$=2-1 $^{13}$CO in July 2006. Weather
conditions were good to excellent. System temperatures including the
sky were 160 K to 240 K for the $J$=1-0 transitions and 400 - 500 K for
the $J$=2-1 transitions. Beam sizes are $21"$-$22"$ and 11" at 
110-115 GHz and 220-230 GHz corresponding to these
transitions. Main-beam efficiencies were 0.74, 0.73,
0.48, and 0.45 at these four frequencies, respectively. The $J$=2-1
$^{12}$CO and $^{13}$CO lines were also observed with the JCMT 15 m
telescope in June and July 1995, with overall system temperatures
including the sky of 485 and 340 K, respectively. The beam size was
$21"-22"$, and the main-beam efficiency was 0.69. All spectra were
binned to resolutions of 20 km/s. A linear baseline was subtracted,
and the line flux was determined by integrating over the velocity
range V(LSR) = 2800 - 3500 km/s.

The HCN J=(3-2) observations were made with the JCMT in February 2010
using receiver A3 under good weather conditions with system
temperatures of 240 to 310 K; the beam size was $18"$, and we used a
main-beam efficiency of 0.69 at the operating frequency of 265.9
GHz. HCN J=(4-3) was obtained with the HARP array in stare mode
on the JCMT in May 2010.  Weather was excellent, with T(sys) in the
range of 226-240 K. We extracted the line profile from the central
pixel. The beam size was about $13"$ and the main-beam efficiency
about 0.6.  From the observed spectra, line fluxes were recovered in
the same way as for the $^{13}$CO observations.

\subsection{Data reduction}
\label{sec:obs1} 

{\it Herschel SPIRE FTS data:} The data were reduced using version 9.0 of Herschel Interactive Processing
Environment (HIPE). For all extended sources, an aperture correction is
necessary to compensate for the wavelength dependent beam size.
This requires knowledge of the source distribution at SPIRE wavelength. We
approximated the size based on
a high spatial resolution SMA CO J=3-2 map \citep{2008ApJS..178..189W}.

Each SMA map was convolved with a 2-D Gaussian to match the FWHM of the
SPIRE beam sizes (15-42''). We then determine the flux density at the
SPIRE pointing centers as a function of spatial resolution normalized
by the flux density in the largest aperture (42''). The resulting
dependency between normalized flux density and spatial resolution was
then applied to SPIRE's Long Wavelength Spectrometer Array (SLW) and the Short Wavelength Spectrometer Array (SSW) spectra taking the SPIRE beam sizes as
a function of wavelength into account. Finally the SLW and SSW spectra
were coadded flagging the noisy edge channels in both spectra. This
yields a combined spectrum at an effective spatial resolution of 42''
for each source. 

The quality of the aperture correction can easily be evaluated by
comparing the continuum flux densities in the corrected SLW and SSW
spectra in their spectral overlap region. Our
approach effectively removes the 'jump' visible in the continua
between the SLW and SSW spectra at their original spatial resolution, although 
we only present the original observed spectra below.

The ratio of the flux between each convolved SMA map and the flux within
the largest beam size (42'') is the
beam correction factor ($\kappa_S$) where:
\begin{equation}
 F_{corr}=F_{obs}\times\kappa_S
\end{equation}
Thus, all fluxes are normalized to a beam size of 42'' (i.e. 9.8
kpc). The beams for pointings B and C significantly overlap, thus it
is hard to discern any independent measurements from these pointings.
However, pointing A is more isolated.  Although the largest beam does
include some of B and C, most of the beam sizes are completely
independent.

Fluxes were first extracted using FTFitter (https://www.uleth.ca/phy/naylor/index.php?page=ftfitter), a program specifically created to extract line fluxes from Fourier transform spectrographs.  This is an interactive data language (IDL) based graphical user interface that allows the user to fit lines, choose line profiles, fix any line parameter, and extract the flux.
We define a polynomial baseline to fit the continuum and derive the flux from the baseline subtracted spectrum.  
In order to more accurately determine
the amplitude of the line, we fix the FWHM to the expected line width
of $^{12}$CO at each source, using the velocity widths measured by
\citet{2012ApJ...753...46S}.  In the case of very narrow linewidths,
more narrow than the instrumental resolution (J= 4-3 through 8-7 for
Arp 299 C), we do not fix the FWHM but fit the lines as an unresolved
profile.  We use an error of 30\% for our fluxes, which encompasses
our dominant sources of error. Specifically, the uncertainty of the beam size correction using SMA CO J=3-2 map is $\sim$20\%. The error of the absolute
calibration uncertainty for staring-mode SPIRE FTS
observations is an additional 6\% \citep{2014arXiv1403.1107S}.  We also have some
uncertainty in the definition of the baseline and flux extractions,
since we use an unresolved or Gaussian profile for all emission lines, accumulating to $\sim5\%$.

{\it Herschel SPIRE photometry data:} SPIRE maps were reduced using HIPE 10.3.0 \citep{2010ASPC..434..139O} and the SPIRE calibration tree v.10.1. A baseline algorithm \citep{2010A&A...518L..65B} was applied to every scan of the maps in order to correct for offsets between the detector timelines and remove residual baseline signals. Finally, the maps were created using a naive mapping projection. The global fluxes for Arp 299 are measured to be 
21.8, 7.34 and 2.37 Jy for 250, 350 and 500 $\mu$m respectively. For the errors in the SPIRE photometry we adopted a 15\% calibration uncertainty for extended emission; 
(SPIRE Observers Manual, v2.4, 2011).

{\it Herschel PACS data:} The data were processed and calibrated using
HIPE version 10.0 and the pipeline for range spectroscopy. The object was
centered on the 9.4\arcsec central spaxel of the 5 by 5 PACS
array. Little flux is seen outside this central spaxel, and therefore
the fluxes are extracted from the central spaxel and referenced to a
point source. We use a 3 by 3 spaxel correction for extended sources and small pointing
offsets. We used SPLAT as part of the STARLINK software package to
subtract baseline, and determine the peak flux, full-width-half
maximum (FHWMs), integrated flux, and its uncertainty for the CO
lines. To find the integrated flux and uncertainty, we fit a Gaussian profile to the line and integrate the Gaussian.

{\it Ground Based Data:} In the reduction of the line profiles observed with IRAM and JCMT, we
used the CLASS package. The JCMT data were retrieved with the SpecX
package and turned into FITS files which were subsequently imported
into CLASS. The IRAM profiles were immediately available in CLASS
format.  For all line profiles, second-order baselines were
subtracted. Line fluxes were determined both by Gaussian fitting, and
by straightforward summing over a sufficiently wide velocity
interval. Both methods yielded nearly identical results.  We then scaled the $^{13}$CO J=1-0 and J=2-1 up
to the 42'' beamsize using the same method described in Section~\ref{sec:obs1} for the SPIRE FTS observations.

\section{Results}
\label{sec:res}
Here we present the spectral profiles and line fluxes for the SPIRE FTS spectra and the PACS observations. 

\subsection{SPIRE FTS line fluxes}
\begin{figure*}
\centering
\includegraphics[width=17cm]{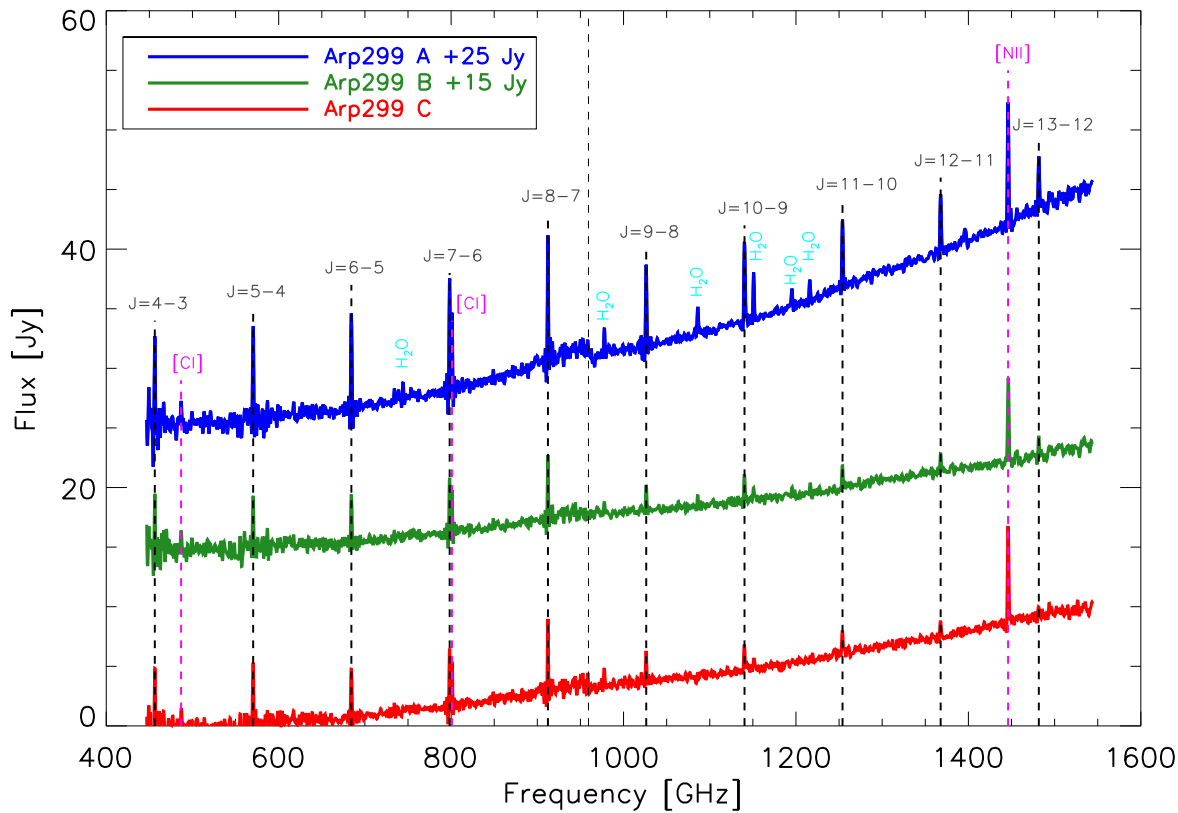}
\caption{SPIRE spectra of Arp 299 A (blue), B (green) and C (red).  Source A is offset by 25 Jy, while Source B is offset by 15 Jy.  Each bright line is identified, CO lines in black, atomic lines in magenta, and H$_2$O lines in cyan. The 
atomic lines and H$_2$O transitions will be discussed in a different paper.}
\label{fig:spectra}
\end{figure*}

The $^{12}$CO transitions are visible from J=4-3 to
J=13-12.  There were also strong detections of [NII] at 1437 GHz and
[CI] at 484 GHz and 796 GHz in all three spectra.  We detect 7 strong
water emission lines, they are most prominent in Source A and
become weaker or undetectable in Sources B and C. The lines are
labeled in Figure~\ref{fig:spectra}, $^{12}$CO in black, H$_2$O in blue, and atomic
lines in magenta. 
As seen in this plot there is a discontinuity between the high
and low frequency modes of the spectrometer.  This discontinuity is
due to the different apertures used by the high and low frequency arrays combined 
with the fact that the object is not a point source.  A scaling factor
($\kappa_S$) for each wavelength is calculated using the method
described in Section~\ref{sec:obs1}, and displayed in
Table~\ref{tab:flux}.

\begin{table*}
\caption{Observed line fluxes corrected for beam size using correction factors ($\kappa_S$, Section~2.2).  Also, the errors on all derived fluxes are 30\% as explained in the text.  Fluxes from ground-based
observations found in the literature are also presented.}
 \begin{tabular}{|l||c|c|c|c|c|c|c|}
\hline
Line &  $\kappa_{S_{A}}$ & Flux Arp 299 A & $\kappa_{S_{B}}$ &Flux Arp 299 B &$\kappa_{S_{C}}$ & Flux Arp 299 C \\
     & 32.8'' & [10$^{-17}$ W m$^{-2}$]& 32.8'' & [10$^{-17}$ W m$^{-2}$]& 32.8''  &[10$^{-17}$ W m$^{-2}$]\\
\hline     
$^{12}$CO 4-3&1.01&8.89&1.02&5.28&1.02&5.88 \\
$^{12}$CO 5-4&1.10&10.8&1.26&6.75&1.24&7.45 \\
$^{12}$CO 6-5&1.14&12.5&1.40&7.18&1.38&7.02 \\
$^{12}$CO 7-6&1.07&13.0&1.19&6.39&1.18&6.29\\
$^{12}$CO 8-7&1.05&14.2&1.12&6.82&1.12&7.31\\
$^{12}$CO 9-8&1.30&13.4&1.39&3.86&1.30&3.97\\
$^{12}$CO 10-9&1.33&14.5&1.51 &4.80&1.45&3.65 \\
$^{12}$CO 11-10&1.33&13.2&1.53&3.84&1.47&3.37\\
$^{12}$CO 12-11&1.34&11.4&1.57&2.72&1.53&2.29\\
$^{12}$CO 13-12&1.35&10.9&1.59&3.45&1.55&1.66\\
$[CI] ^3P_1-^3P_0$&1.04 &2.56&1.09&2.17&1.08&1.94\\
$[CI] ^3P_2-^3P_1$& 1.07&8.46&1.19&4.61&1.18&4.49\\
$[NII] ^3P_1-^3P_0$&1.35 &25.6&1.59&11.1&1.55&5.86 \\
\hline
\hline
$^{12}$CO 1-0&--&0.29&--&0.08\tablefootmark{a}&--&0.01\tablefootmark{a}\\
$^{12}$CO 2-1&--&1.29&--&0.76\tablefootmark{a}&--&0.73\tablefootmark{a}\\
$^{12}$CO 3-2\tablefootmark{a}&--&5.09&--&2.10&--&3.04\\
\hline
$^{13}$CO 1-0&--&0.01&--&--&--&--\\
$^{13}$CO 2-1&--&0.15&--&--&--&--\\
HCN 1-0\tablefootmark{b}&--&0.003&--&--&--&--\\
HCN 3-2&2.52&0.04&--&--&--&--\\ 
HCN 4-3&3.77&0.03&--&--&--&--\\ 
\hline
 \end{tabular}
\tablefoot{ \\
 \tablefoottext{a}{Determined from the maps presented in \citet{2012ApJ...753...46S}.} \\
 \tablefoottext{b}{From \citet{2006PASJ...58..813I}.} \\
 }
\label{tab:flux}
\end{table*}

\subsection{PACS line fluxes}
\label{sec:results}
\begin{figure}
\resizebox{\hsize}{!}{\includegraphics{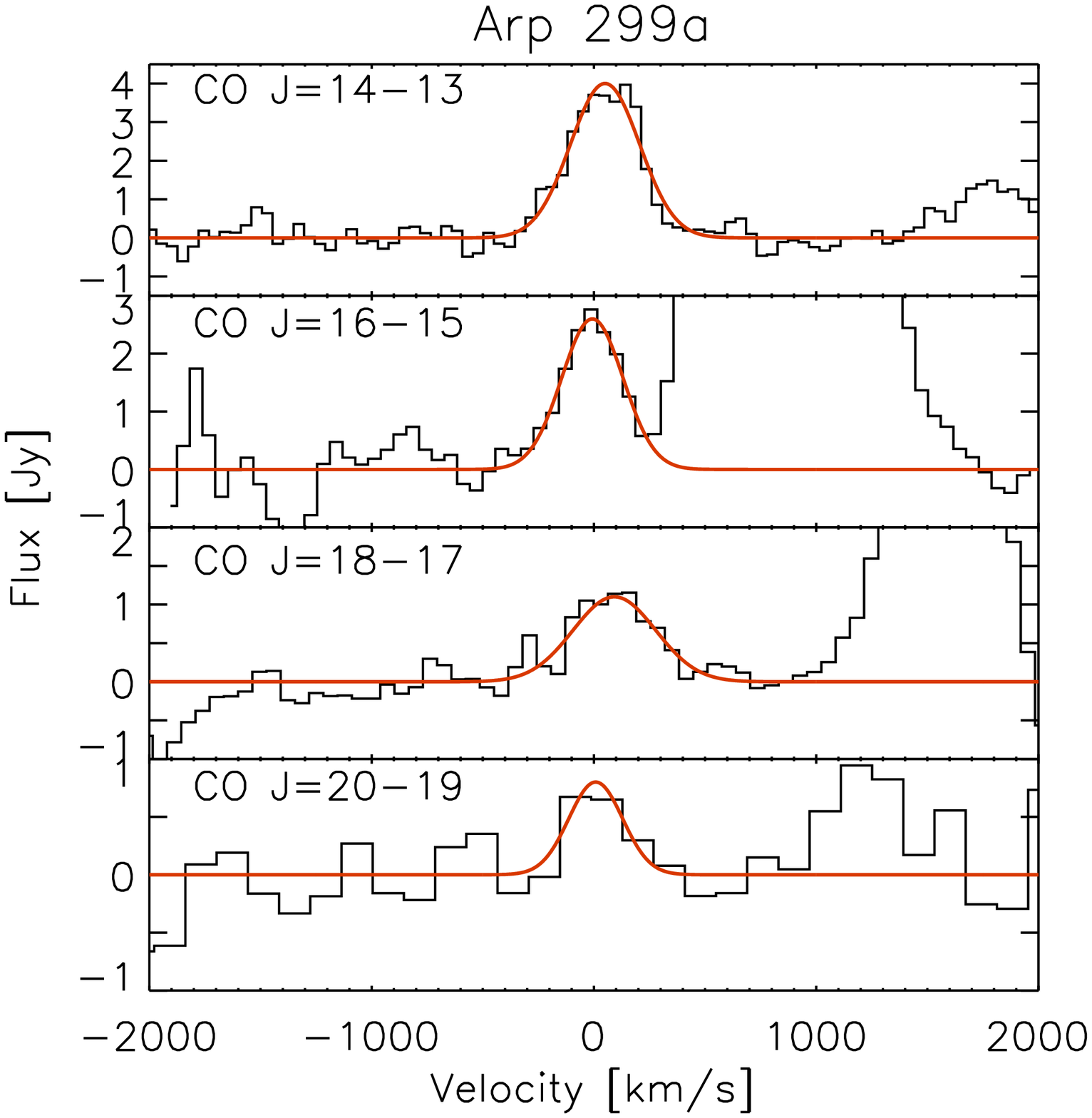}}
\caption{PACS spectra of Arp 299 A, showing the detections for CO J $=
  14-13$, $16-15$, $18-17$, and $20-19$. The J$=20-19$ transition is only a 2$\sigma$ detection. The 
  line to the right of the J=$16-15$ transition, around 1000 km/s is the OH doublet ($\Pi_{1/2}-\Pi_{1/2}$ 3/2-1/2 163.124-163.397 $\mu$m), and the line to the right of the J=18-17 transition is
  the [OI] 145$\mu$m line.  The velocity scale is calculated for a redshift of $z=0.01030$ and the channel spacings are 47, 57, 64, and 140 km/s
  for the J $=14-13$, $16-15$, $18-17$, and $20-19$ respectively.}
\label{PACS_SPEC}
\end{figure}

The PACS CO $J = 14-13$, $16-15$, $18-17$, and $20-19$ line
detections are shown in Fig. \ref{PACS_SPEC} and their peak flux,
FWHM, and integrated fluxes are listed in Table
\ref{PACS_LINE_PARAMS}. We also would like to 
note that the CO $J=20-19$ transition is only detected at 2$\sigma$. The CO $J=22-21$, $J=24-23$ and $28-27$ were
not detected, and for these lines we determined an upper limit.

\begin{table*}
  \caption[]{PACS CO observations.  }
\label{PACS_LINE_PARAMS}    
\begin{center}
\begin{tabular}{l c c c c c c } 
\vspace{-5.0mm}    \\
\hline\hline   
\noalign{\smallskip}
Transition   &  $\lambda_{\rm rest}$ & Peak  &  $\Delta V_c$\tablefootmark{a}  & FWHM  & $S_{\rm line}$     & $S_{\rm line, corr}$ \\
             &   [$\mu$m]         &  [Jy] &  [km/s] & [km/s] & [10$^{-17}$ W m$^{-2}$]& [10$^{-17}$ W m$^{-2}$]\\
\noalign{\smallskip}
\hline
\noalign{\smallskip}
\multicolumn{7}{c}{Arp 299 A} \\
\noalign{\smallskip}
\hline               
\noalign{\smallskip}
CO $J=14-13$ & 185.999    & $4.0\pm 0.2$ & $50\pm7$   & $367\pm17$  & $8.3\pm0.5$ & $11.2\pm0.7$ \\
CO $J=16-15$ & 162.812    & $2.6\pm 0.1$ & $-7\pm6$   & $326\pm16$  & $5.6\pm0.3$  & $7.5\pm0.5$ \\
CO $J=18-17$ & 144.784    & $1.1\pm 0.1$ & $92\pm21$  & $438\pm50$  & $3.6\pm0.5$  & $4.8\pm0.7$  \\
CO $J=20-19$ & 130.369    & $0.8\pm 0.2$ & $8\pm36$   & $285\pm87$  & $1.8\pm0.7$  & $2.4\pm0.9$  \\
CO $J=22-21$ & 118.581    &              &            &             & $< 2.3$     & $<3.1$ \\
CO $J=24-23$ & 108.763    &              &            &             & $< 2.6$     & $< 3.5$ \\
CO $J=28-27$ & 93.3491    &              &            &             & $< 1.7$     & $< 2.3$ \\
\noalign{\smallskip}
\hline               
\noalign{\smallskip}
\end{tabular}
\tablefoot{ \\
 \tablefoottext{a}{$\Delta V_c$ is the distance in km/s away from the central wavelength of the line.} \\
 \tablefoottext{b}{From \citet{2006PASJ...58..813I}.} \\
 }
\end{center}
\end{table*}

\section{Comparison between Arp 299 A and B+C}
\label{sec:comp}
In this section, we perform a comparison of Arp 299 A, B, and C using only 
the SPIRE FTS fluxes to determine what differences are observed from $^{12}$CO alone. 
The most notable aspect of the spectra presented in Figure~\ref{fig:spectra}
is that the high-J CO lines of Arp 299 A are distinctly brighter than
those of Arp 299 B and C.  It is clear simply from inspecting the
spectra that the molecular gas in Arp 299 A is more excited than that of 
Arp 299 B and C.  

\begin{figure}
\resizebox{\hsize}{!}{\includegraphics{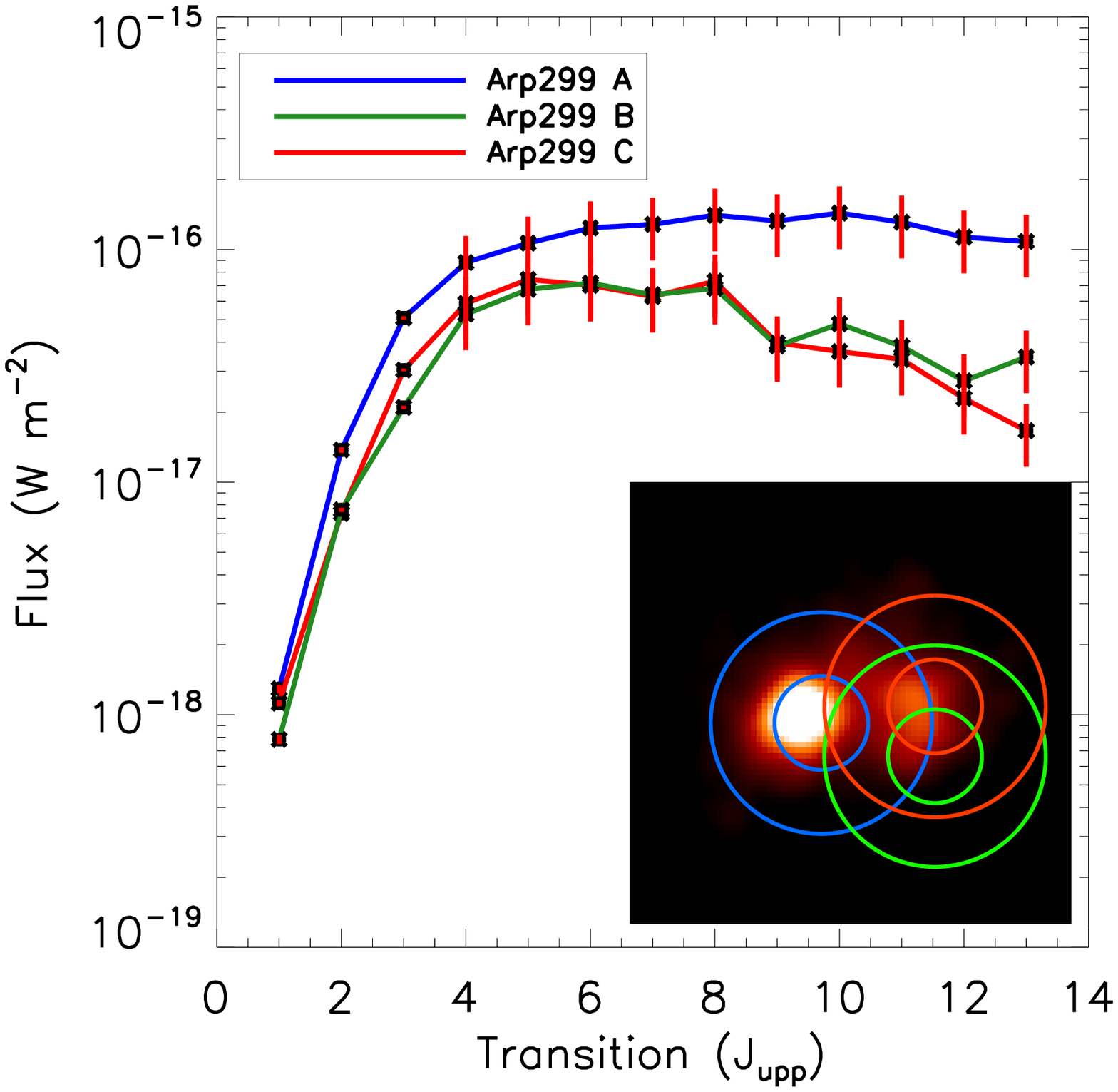}}
\caption{CO excitation ladders of Arp 299 A (blue), B (green), and C (red).  Intensities are in W m$^{-2}$. The 
inset in the bottom right corner shows the three SPIRE beam FWHMs for Arp 299 A, B, and C overplotted on a 
SCUBA 450 $\mu$m archival image.  The smaller circle 
represents the smallest beam FWHM ($\sim$17'') and the larger circle represents the largest beam FWHM ($\sim$42'').}
\label{fig:coladder}
\end{figure}

For each spectrum (A, B, and C) we can create a spectral line energy
distribution or 'CO ladder', which plots the intensity of each CO
transition as a function of the upper J number.  This type of diagram
is predicted to be a powerful diagnostic tool as shown by
\citet{2005A&A...436..397M} and \citet{2007A&A...461..793M}, where
models show that these CO ladders have very different shapes depending
on the type of excitation (i.e. photon dominated region, PDR or X-ray
dominated region, XDR) as well as density and radiation environment.
The three CO ladders for Source A, B, and C are plotted on top of each
other in Figure~\ref{fig:coladder}.  For context, their smallest and
largest beam sizes are plotted over a SCUBA 450 $\mu$m image, showing
the overlap between the Arp 299 B and C pointings.  This overlap is
also apparent in the CO ladders, the two ladders follow the same shape
and intensity, meaning they are essentially an averaged observation of
both Arp 299 B and C.  Because of this, we only use the averaged
values for Arp 299 B and C from here on.  Although we cannot discern
anything independent about Arp 299 B and C, it is immediately apparent
that Arp 299 A has a very differnt CO ladder.  Arp 299 A flattens in
intensity with increasing transitions, while Arp 299 B and C both show
a turnover in their ladders at J$_{upp}$=5.  This indicates clearly that there is more
warm CO in Arp 299 A than in B+C and we expect to see this reflected in the following PDR analysis.

\subsection{Basic PDR analysis}
Since Arp 299 is a LIRG with a high star formation rate, there must be a high density of OB stars and thus a high UV energy density.  
Through photoelectric heating and FUV pumping of H$_2$, the 
FUV photons heat the outer layers (A$_V$<5) of molecular clouds.  This area of the molecular cloud is the PDR, and is responsible
for warm molecular gas emission.  The thermal state of PDRs is determined by processes such as photo-electric heating; heating by pumping of H$_2$ followed by collisional de-excitation; 
heating by cosmic rays; [OI] and [CII] fine-structure line cooling; and CO, H$_2$O, H$_2$, and OH molecular cooling. The ionization degree of the gas is driven by FUV
photo-ionization, and counteracted by recombination and charge transfer reactions with metals and PAHs. The ionization degree is at most $x_{\rm e} \sim 10^{-4}$ outside of the fully ionized zone. 
The chemistry exhibits two fundamental transitions, H to H$_2$ and C$^+$ to C to CO.  
Using PDR models \citep{2005A&A...436..397M,2012A&A...542A..65K} that solve for chemistry and thermal balance throughout
the layers of the PDR, we use the predictions of the $^{12}$CO emission as a function
of density, radiation environment ($G$, in units of the Habing radiation field G$_0$=1.6$\times10^{-3}$ erg cm$^{-2}$ s$^{-1}$), 
and column density. We use an isotopic abundance ratio 
of 80 for $^{12}$CO/$^{13}$CO, since our observed $^{12}$CO/$^{13}$CO J=1-0 intensity ratio is $\sim24$, which is 
common in (U)LIRGs \citep{1997ApJ...475L.107A}. \citet{2012A&A...541A...4G} find an isotope ratio
around 100 for the prominent starburst Arp 220, which is similar to that of Mrk 231 \citep{2014A&A...565A...3H}.  However, for a less powerful starburst, such as 
NGC 253, the isotope ratio was measured to be 40 \citep{2014A&A...565A...3H}.  Since Arp 299 is a moderate starburst, an estimate of 80 is reasonable.  The density profile is constant and the Habing field 
is parameterized in units of G$_0$ from photons between 6 eV and 13.6 eV. We perform an unbiased fitting of the models to the CO ladder, employing an
automated $\chi^2$ fitting routine, described in detail in \citet{2014arXiv1401.4924R}.  This routine allows for up to 3 different ISM phases
where we define the total model as:
\begin{equation}
\label{eq:tot_mod}
Model=\Omega_IPDR_I+\Omega_{II}PDR_{II}+\Omega_{III}PDR_{III}
\end{equation}
\noindent where
$PDR_I$, $PDR_{II}$, and $PDR_{III}$ are the distinct contributions of the three PDR
models.  $\Omega_I$,
$\Omega_{II}$, and $\Omega_{III}$ represent the respective filling
factors of each ISM phase.  Filling factors traditionally 
represent how much of the beam is filled, so they only range from 0 to 1.  However, this assumes that
these clouds do not overlap in velocity, which we allow for.  Thus, $\Omega$ is not only a beam filling factor, but 
also allows for an overlap in velocity, which accounts for it being slightly greater than one.  

We perform a modified Pearson's $\chi^2$ minimized fit for $^{12}$CO and $^{13}$CO simultaneously, where the modified Pearson's $\chi^2$ is:
\begin{equation}
\chi^2_{mol}=\frac{\sum_{i=1}^{N_{data}} \frac{(obs_i-model_i)^2}{model_i}}{N_{data}}
\label{eq:chi}
\end{equation}
We define $\chi^2_{mol}$ as the modified Pearson's $\chi^2$ for a specific molecule.  The total $\chi^2$ is the sum of the $\chi^2_{mol}$ terms for each molecule.  The numerator of this equation is the traditional Pearson's $\chi^2$, then in the denominator we divide
by the total number of transitions in each respective molecule, essentially yielding an average $\chi^2$ for $^{12}$CO and $^{13}$CO separately.  In Section 5, we refer to the total $\chi^2$ as being the sum of Eq.~\ref{eq:chi} for all molecules; $^{12}$CO, $^{13}$CO, and HCN. 

\begin{figure*}
 \centering
  \includegraphics[width=8cm]{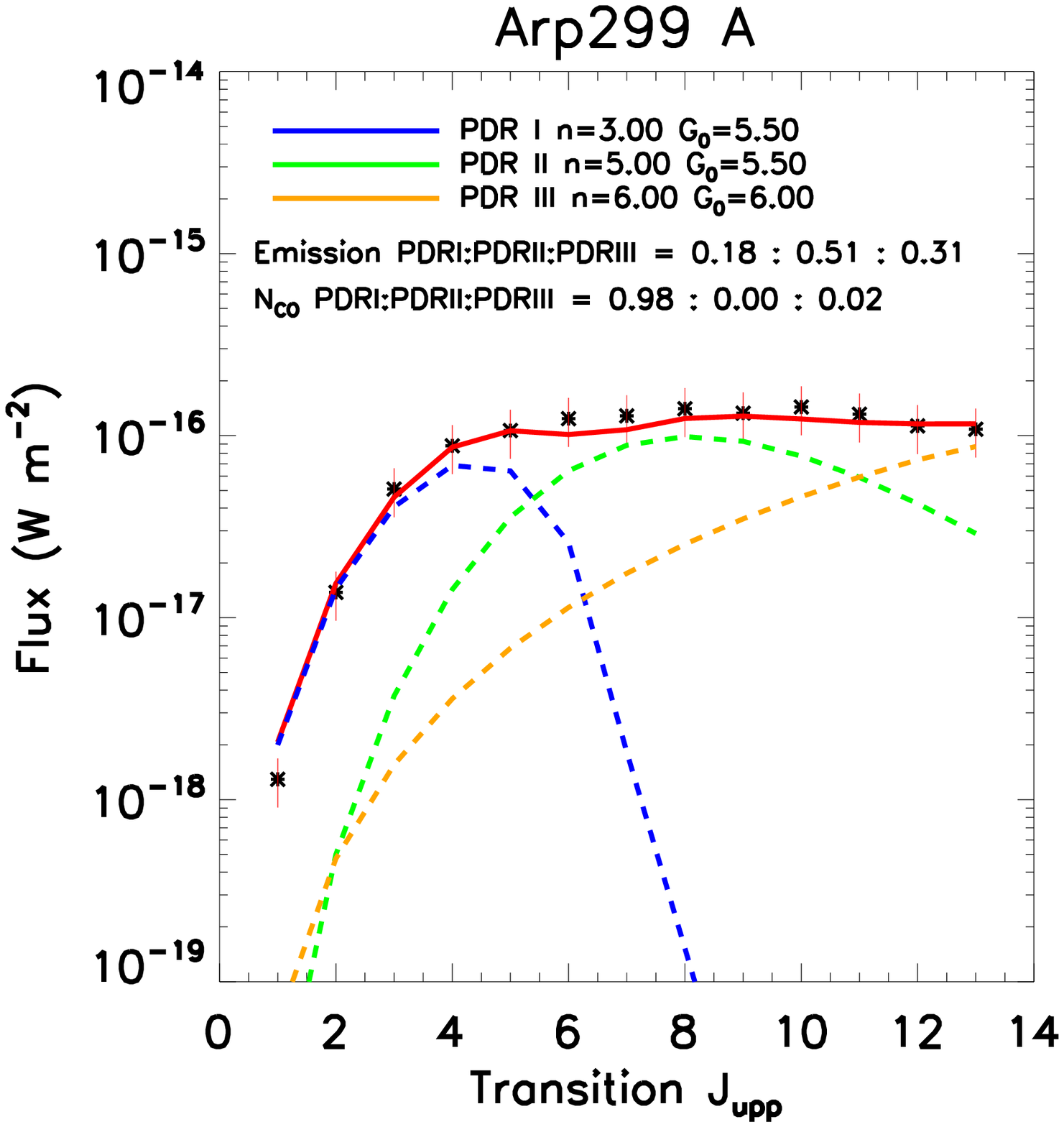}
  \includegraphics[width=8cm]{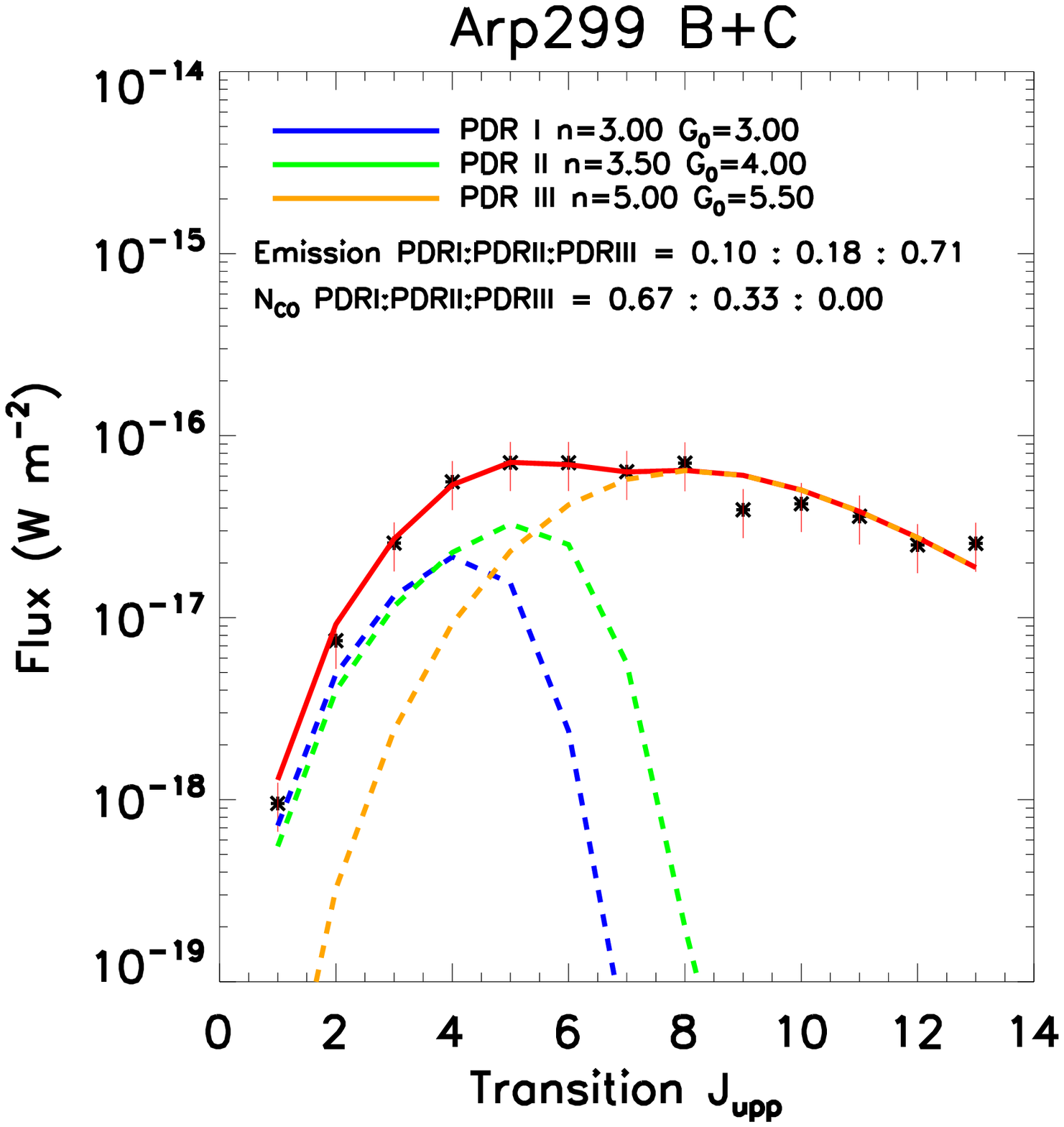}
  \caption{Best fit ($\chi^2$ minimized) PDR models simulating the $^{12}$CO emission for Arp 299 A (left)and B$+$C (right) using three ISM phases.}  The red line is the sum of the three phase 
  models, the black asterisks are the data points with error bars, and the blue, green, and orange lines represent 
  the independent PDR models for each phases. The model density, temperature and column density are shown in the legend along with the relative contribution of 
  each phase in terms of emission and column density.
 \label{fig:pdrfit}
\end{figure*}

Using this equation, we calculate the
$\chi^2$ for every combination of 3 models and filling factors.  In this way, we cannot only see which models make the best
fit, but we can also see the $\chi^2$ values for all the other model
combinations.  This allows us to understand the level of degeneracy
inherent to the models and understand the limitations of this method.
In Figure~\ref{fig:pdrfit}, we show the best fitting models for Arp 299 A
and Arp 299 B+C.  We also calculate the relative contribution of each
independent model to the overall CO ladder intensity in terms of
emission and CO column density.

\begin{figure*}
\centering
\includegraphics[width=6cm]{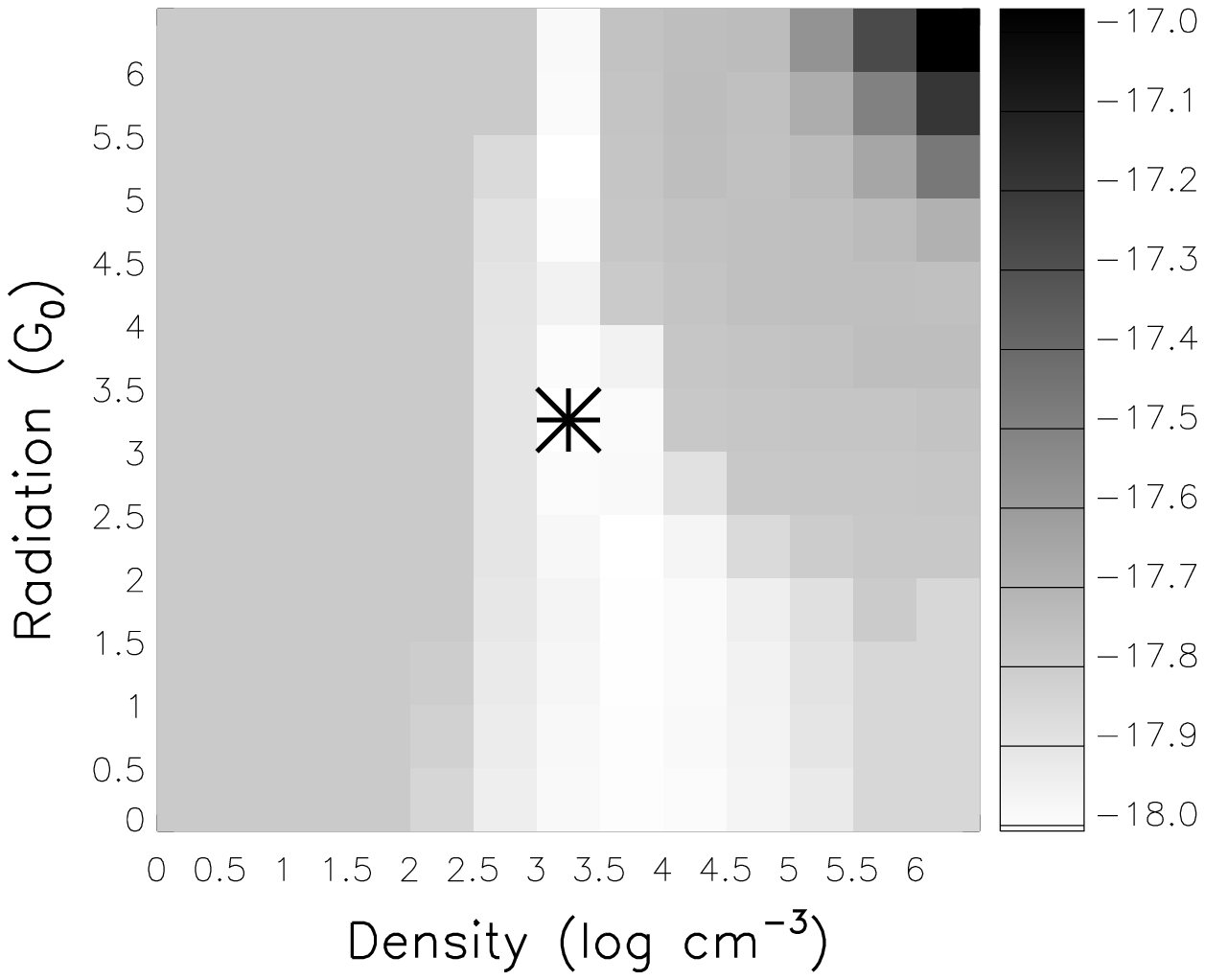}
\includegraphics[width=6cm]{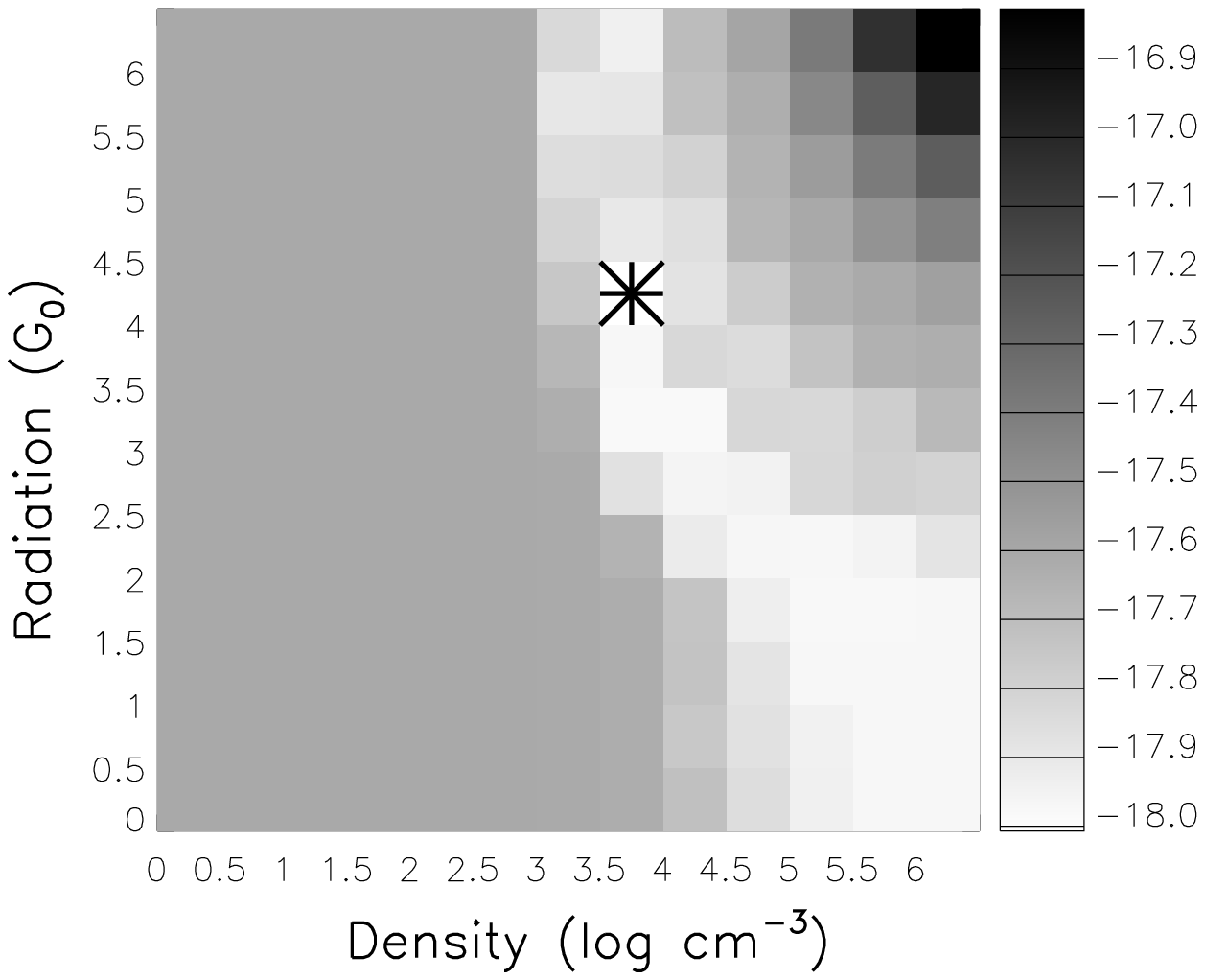}
\includegraphics[width=6cm]{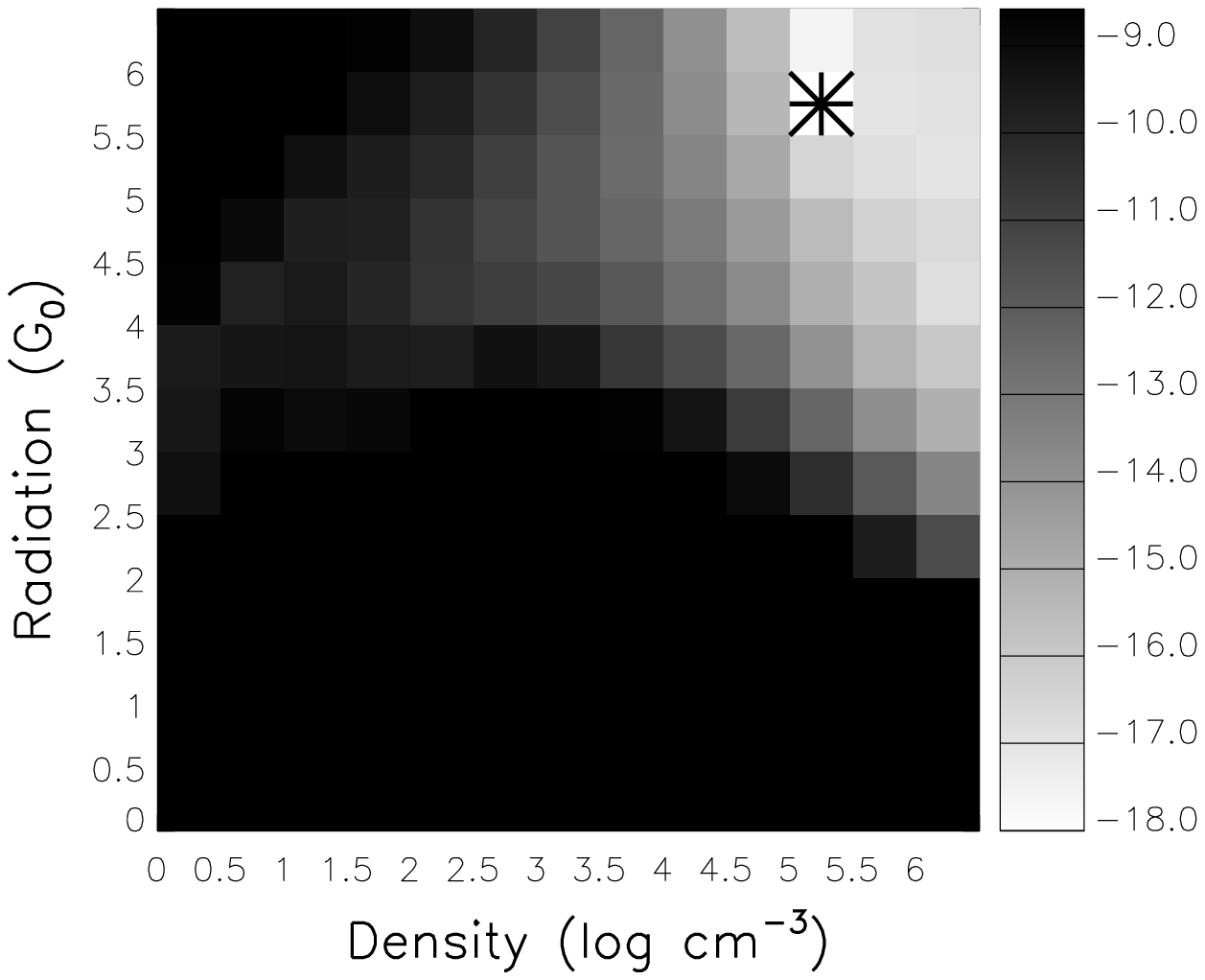}
\caption{Degeneracy plots for Arp 299 B+C.  Each plot represents the full grid of models for each phase in Figure~\ref{fig:pdrfit}, 
PDR I (left), PDR II (center), PDR III(right). The asterisk represents the model with the lowest $\chi^2$, which are also plotted in 
Figure ~\ref{fig:pdrfit}.  The gray scale of each small square indicates the associated log($\chi^2$) value for that particular model, white 
being the best fits and black being the worst. The $\chi^2$ value is defined in Equation~\ref{eq:chi} and shown in log gray scale.}
\label{fig:degenC}
\end{figure*}

One aspect of these fits is that each of the CO ladders needs a
minimum of three ISM phases to be fit well.  In addition, the lowest J
transitions are fit with a relatively low density and low $G$ PDR, the
middle phase is a medium density and medium $G$ PDR, and finally the
highest J transitions can only be fit by extreme PDRs, which makes up
a negligible percent of the CO column density, but over 30\% of the
total CO emission in the case of Arp 299 A and over 60\% of the total
CO emission in Apr299 B+C.  In Figure~\ref{fig:degenC}, we display the
degeneracy plots for Arp 299 B+C.  These plots are only a slice of the full degeneracy cube, held
at the best fit column densities.  They are a representative example of
the degeneracy plots of the other fits and share similar
characteristics.  In the left panel, we show the degeneracy plot for
the first ISM phase (PDR I).  Each small square represents a different
model with a particular density and radiation.  The color represents
the $\chi^2$ value, white being the lowest and black being the
highest.

As seen in Figure~\ref{fig:degenC}, the fits are degenerate.  We do
have a 'best fit', designated with an asterisk, but especially in the case of PDR I, there are a
wide range of models that would fit almost as well as the selected
model.  We can only constrain density to $n=10^{2.5}-10^{5.5}$ cm$^{-3}$ and the radiation field in unconstrained. Even though 
the fits are degenerate, it is clear that each
phase has a specific and independent range of parameter space for
which there is a good fit. Each ISM phase has a trade off between
radiation and density, but each cover a different range of values.  For instance, PDR I ranges in density from 
$n=10^{2.5}-10^{5.5}$ cm$^{-3}$, while PDR II ranges from $n=10^{3}-10^{6}$ cm$^{-3}$ and PDR III ranges from 
$n=10^{5.0}-10^{6}$ cm$^{-3}$.  The radiation field strength is not as well constrained and 
varies inversely to the density.  However without 
more information, we cannot break the observed degeneracies. 

Since we have ancillary data for Arp 299 A, and since we cannot 
 separate the contributions from Arp 299 B and C, the following
discussion will focus on Arp 299 A.

\section{A case study: Arp 299 A}
\label{sec:case}
Using the PACS high-J $^{12}$CO as well as the JCMT $^{13}$CO, and HCN observations, we determine if Arp 299 A can
be heated purely through UV heating or if additional heating sources are necessary.  We can use the PACS observations presented in Section~\ref{sec:results} to extend
the SPIRE CO ladder from J$_{upp}$=13 to J$_{up}=20$.  We then add the observations of $^{13}$CO J=1-0 and J=2-1 
to constrain column density and observations of HCN J=1-0, J=3-2, and J=4-3 to constrain the high density components.  In addition, we extract fluxes from the SPIRE Photometry maps and combine them with observations from the literature in order
to perform an SED analysis of the dust to help further disentangle UV from other heating sources.

\subsection{The low-excitation phase}
\begin{figure}
\resizebox{\hsize}{!}{\includegraphics{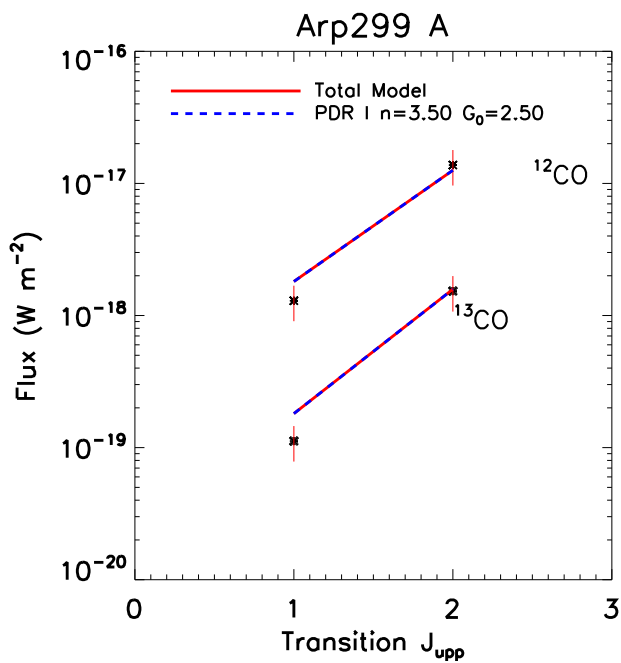}}
\caption{Best fit model of $^{13}$CO J<2 and $^{12}$CO J<2. The red line is the total model, the black asterisks are the data points
with error bars, and the blue dotted line represents 
the first ISM phase, represented by a PDR model. The model density and radiation strength are shown in the legend.}
\label{fig:13}
\end{figure}

Before we blindly fit the full grid of models to our observations, we can constrain the first ISM phase, responsible
for the low-J CO lines.  Since $^{13}$CO is optically thin, the ratio of $^{13}$CO to $^{12}$CO constrains the optical
depth, and in turn the column density.  We have observations of $^{13}$CO 1-0 and 2-1 from the JCMT, as presented in
Section~\ref{sec:obs}.  We can assume that the  $^{13}$CO 1-0 and 2-1 lines arise from the same ISM phase that is responsible
for the first few transitions of $^{12}$CO, and can run the automated fitting routine on the low-J transitions alone.  The 
best fit is displayed in Figure~\ref{fig:13}.  Since often, most of the $^{12}$CO is in low rotational states, 
we can help constrain the mass of the whole system by finding the mass of the low-excitation ISM phase.  
In this phase, we find a mass of 2$\times10^9$ M$_\odot$ (Eq.~\ref{eq:col}), which represents $\sim$66\% of the total molecular gas mass.

\subsection{Full PDR analysis}
\label{sec:pdr}
\begin{figure*}
\centering
\includegraphics[width=17cm]{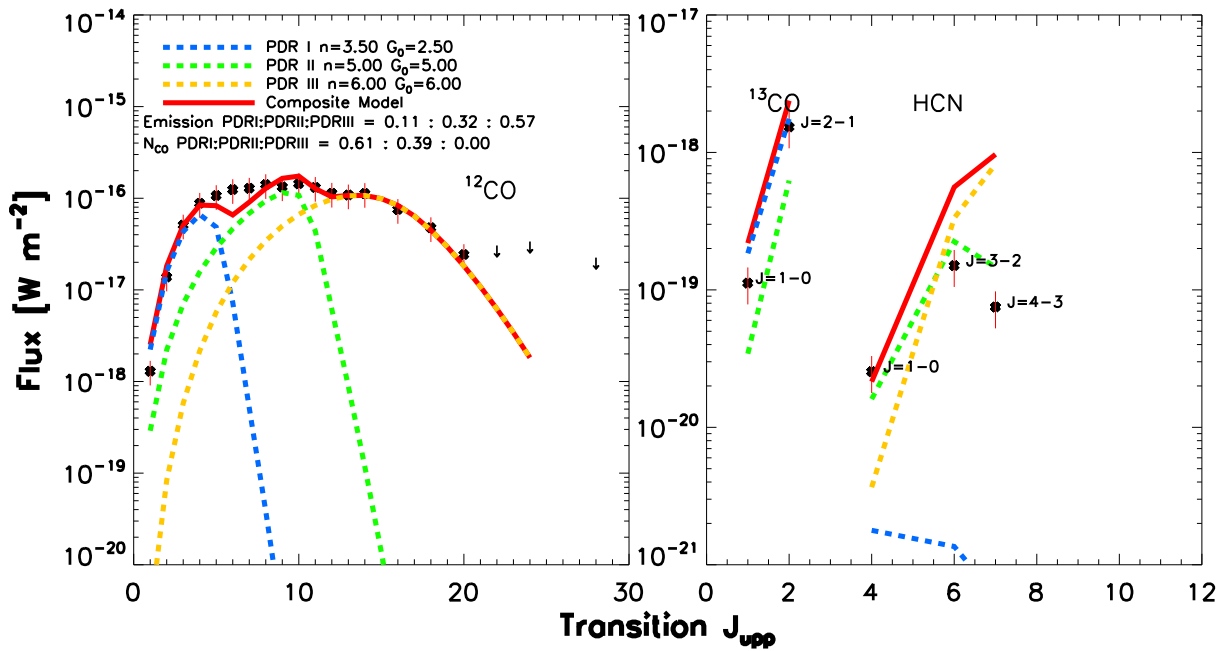}
\caption{\textbf{Left panel:} A $\chi^2$ minimized fit of the CO ladder of Arp 299 A from J=1-0 through 24-23.  The fit was constrained to 
three PDR models, displayed in blue, green, and yellow.  The black points are the observed CO fluxes and the red line is the total 
model fit.  \textbf{Right panel:} The same models as the left panel, but the fits for the $^{13}$CO and HCN fluxes.  The HCN model lies
far above the observed J=3-2 and 4-3 HCN fluxes.}
\label{fig:highj}
\end{figure*}

Now that we have constrained the first ISM phase, we can include all the available data to constrain the other
ISM phases.  We will include $^{12}$CO observations from PACS including J=14-13 through J=20-19 to constrain the
CO ladder turn-over point, HCN 1-0, 3-2, and 4-3, to constrain the properties of the high density gas, and the SPIRE photometry observations
to estimate the dust temperature. 
With all the available line fluxes, 
we can first fit the full CO  and HCN ladders of Arp 299 A using pure PDR models. 
In Figure~\ref{fig:highj}, we display the $^{12}$CO and $^{13}$CO ladders from J=1-0 through
28-27 and J=1-0 to 2-1 respectively, along with the $\chi^2$ minimized fits.  We also calculate the relative contribution
of each independent model to the overall CO ladder in terms of
luminosity and CO column density.

The parameters of the fits are given in Table~\ref{tab:pdr}.  We have estimated the masses of each 
ISM phase using the equation from \citet{2014arXiv1401.4924R}:

\begin{equation}
\label{eq:col}
M_{H_2}=\sum_{i}^{n}\frac{\Omega_i N_{H2,i}A_{beam}m_{H2}}{M_\odot}
\end{equation}  

\noindent where $N_{H2}$ is the H$_2$ column density in cm$^{-2}$
which is consistently calculated in the PDR models, $A_{beam}$ is the
beam area in cmX
$^2$, and $m_{H2}$ is the mass of a hydrogen molecule.

We estimate the relative contributions of column density and emission to the total $^{12}$CO ladder.  In order
to estimate the relative contribution of emission, we use the following equation:
\begin{equation}
C_{em}=\rm\frac{\sum_{i=1}^{13} CO_{mod,i}}{CO_{tot}}
 \label{eq:cem}
\end{equation}
\noindent where $^{12}$CO$_{mod}$ is the summed flux from the modeled CO transitions from J=1-0 to J=13-12 of a specific PDR model and 
$^{12}$CO$_{tot}$ is the total model flux, defined in Eq.~\ref{eq:tot_mod}.  We use the same method for calculating 
the contribution of column density, except we compare the column density of each PDR model to the total column density. 

\begin{table*}
\caption{Model parameters for the three molecular gas ISM phases using only PDR heating.}
 \begin{tabular}{|l||c|c|c|c|c|c|c|c|}
\hline
Component &  Density log(n$_{H}$) & log($G$) & log(N$_{CO}$) &log(N$_{H2}$)&  $\Omega$\tablefootmark{a} & C$_{em}$\tablefootmark{b} & C$_{N_{CO}}$\tablefootmark{c} & Mass$_{N_{H2}}$\tablefootmark{d} \\
          &  log[cm$^{-3}$]  &  G$_0$  & log[cm$^{-2}$] &log[cm$^{-2}$]&  && &M$_\odot$ \\
\hline
\multicolumn{9}{|c|}{ M$_{tot}$: $2\times10^9$ M$_\odot$ } \\
\hline
PDR I    & 3.5 & 2.5 & 17.1& 21.5&1.2 & 0.11 & 0.61 & $2\times10^9$\\
PDR II  & 5.0 & 5.0 & 18.2&21.9 &0.06 & 0.32 & 0.39 & $3\times10^8$\\
PDR III & 6.0 & 6.0 & 16.7 &21.2  &0.006 & 0.57 &$<0.01$ &$6\times10^6$\\
\hline
 \end{tabular}
  \tablefoot{ \\
 \tablefoottext{a}{$\Omega$ is the beam filling factor for each ISM phase.} \\
 \tablefoottext{b}{C$_{em}$ is the fractional contribution of each ISM phase to the emission, as in Eq.~\ref{eq:cem}.} \\
 \tablefoottext{c}{C$_{N_{CO}}$ is the fractional contribution of each ISM phase to the column density.} \\
 \tablefoottext{d}{Mass$_{N_{H2}}$ is the mass of each ISM phase as estimated by the column density using Eq.~\ref{eq:col}.} \\
 }
\label{tab:pdr}
\end{table*}

Three pure PDR models fit the $^{12}$CO well, although the mid-J lines are not all fully reproduced.  The $^{13}$CO is also very well reproduced. We find an H$_2$ mass of $3\times10^9$ M$_\odot$, which matches the mass 
estimates from the literature, 1.8-8.6$\times10^9$ M$_\odot$
 \citep{2012ApJ...753...46S,1987ApJ...312L..35S,1988ApJ...334..613S}. As shown in \citet{2014arXiv1401.4924R}, 
HCN is a good tracer of the excitation mechanism since the relative line ratios of various HCN transitions depend on
excitation mechanism, and in the pure PDR fit, the models fail 
to fit any of the HCN transitions. Note that the red line for HCN lies far above the observed J=3-2 and 4-3 transitions.  Since we cannot reproduce both the CO 
and HCN emission with the same best fit model, this suggests that there is an alternative heating mechanism responsible for heating the dense gas, which is traced by the HCN.  In order
to produce enough CO flux in the high-J transitions, the HCN is overproduced, thus we need a mechanism which selectively heats the high-J CO without heating as much HCN. 

In addition, the only way to reproduce the flux of the high-J CO lines with a PDR is with a density of 10$^6$ cm$^{-3}$ and a radiation flux of 10$^{6}$ G$_0$, 
which is an order of magnitude higher than the most extreme PDRs (i.e. Orion Bar) found in the Milky Way.  In terms of mass, this ISM phase represents about $\sim0.2\%$ 
of the total molecular
gas mass of Arp 299 A (Table~\ref{tab:pdr}).  Since UV photons are even more efficient in heating the dust than the gas (unlike
X-rays, cosmic rays, and mechanical heating), we expect the same percentage of the dust mass to be heated to high temperatures (>200 K). 
Using a combination of three gray bodies, we can fit the SED with a "cool", "warm", and "hot" dust
component (see, e.g., \cite{2010ApJ...715..775P}) aiming to account for the cold cirrus-type dust,
the star formation-heated dust, and an AGN-heated dust respectively. We caution the reader on the simplicity of the
physics underlying this kind of modeling and especially for the emission at mid-infrared wavelengths where
dust is primarily not in thermal equilibrium with the local interstellar radiation field. However this approach provides reasonable estimates for the
average dust temperatures and masses for each component. The dust emissivity is a power law, where $\kappa_{\nu} =
\kappa_{0}^{\beta}$. We assume a value of $\kappa_{0}$ = 0.192 m$^2$ kg$^{?1}$ at 350 $\mu m$
\citep{2003ARA&A..41..241D} and $\beta$ = 2.  {The value $\beta$=2 was adopted as the most suitable for global dust emission SEDs (e.g.,\cite{2001MNRAS.327..697D}). 
It has to be noted though that $\beta$ varies within galaxies (see, e.g., \cite{2014A&A...561A..95T}). } To find the best fit SED our code minimizes the $\chi^2$ function using the Levenberg-Marquardt algorithm \citep{1992drea.book.....B}. Besides the SPIRE data, which were reduced by us, we used the fluxes presented in \citet{2012ApJS..203....9U}. The result of the three component fit is shown in Figure~\ref{fig:highj} (see the figure caption for an explanation of the symbols).

\begin{figure}
\resizebox{\hsize}{!}{\includegraphics{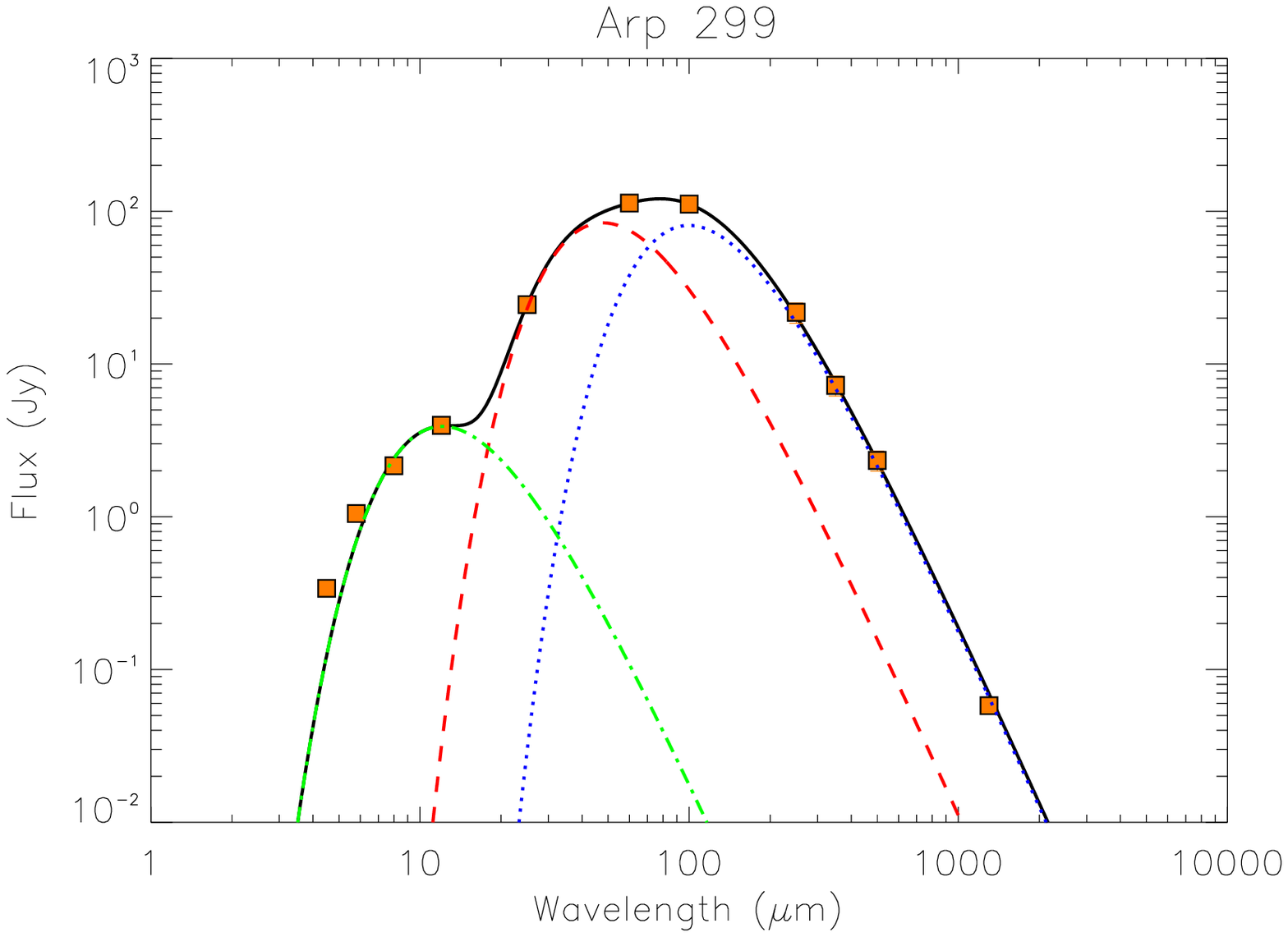}}
\caption{The dust SED of Arp 299. This is a fit with three gray bodies representing
the cold (blue), warm (red) and hot (green) dust components. Observations are plotted with orange squares while the black line is the total SED model fit (Section 4.2).}
\label{fig:dust}
\end{figure}

The temperature and mass of each 
dust component is calculated.  We find a temperature of 29.1$\pm3.5$ K for the "cool" component, 60.6$\pm4.9$ K for the "warm" 
component, and 239.4$\pm22.7$ K for the
"hot" component.  The dust masses are 1.1$\times10^8$ M$_{\odot}$, 2.9$\times10^6$ M$_{\odot}$, and
141 M$_{\odot}$, respectively. We find that the "hot" dust component contains only $\sim 10^{-6}$
of the total dust mass.  This is four orders of magnitude smaller than the 2$\times10^{-2}$ of hot dust
expected from a PDR with the parameters of PDR III (last column of Table~\ref{tab:pdr}).  This along with the poor reproduction of HCN emission 
shows that the third ISM phase cannot be heated purely by UV photons.

\subsection{Additional heating sources}
We can explore alternative heating sources to explain the high-J CO and HCN transitions.  We consider 
cosmic ray heating, X-ray heating, and mechanical heating (shocks and turbulence) as alternative 
heating sources. Cosmic rays can also heat gas in cosmic ray dominated regions (CDRs), which are PDRs with and 
enhanced cosmic ray ionization rate; we employ a typical model for enhanced cosmic ray ionization rate with $\zeta_{CDR}=750\zeta_{gal}$, 3.75$\times10^{-14}$ s$^{-1}$.  Cosmic rays are able
to penetrate into the very centers of molecular clouds, where 
even X-rays have trouble reaching and are typically produced by supernovae.  Similarly, PDRs with additional mechanical 
heating (mPDRs) are due to turbulence in the ISM and may be driven by supernovae, strong stellar winds, jets, or outflows. We parameterize 
the strength of the mechanical heating ($\Gamma_{mech}$) with $\alpha$, which represents the fractional
contribution of mechanical heating in comparison to the total heating at the surface of a pure 
PDR (excluding mechanical heating).  

\begin{figure*}
\centering
\includegraphics[width=17cm]{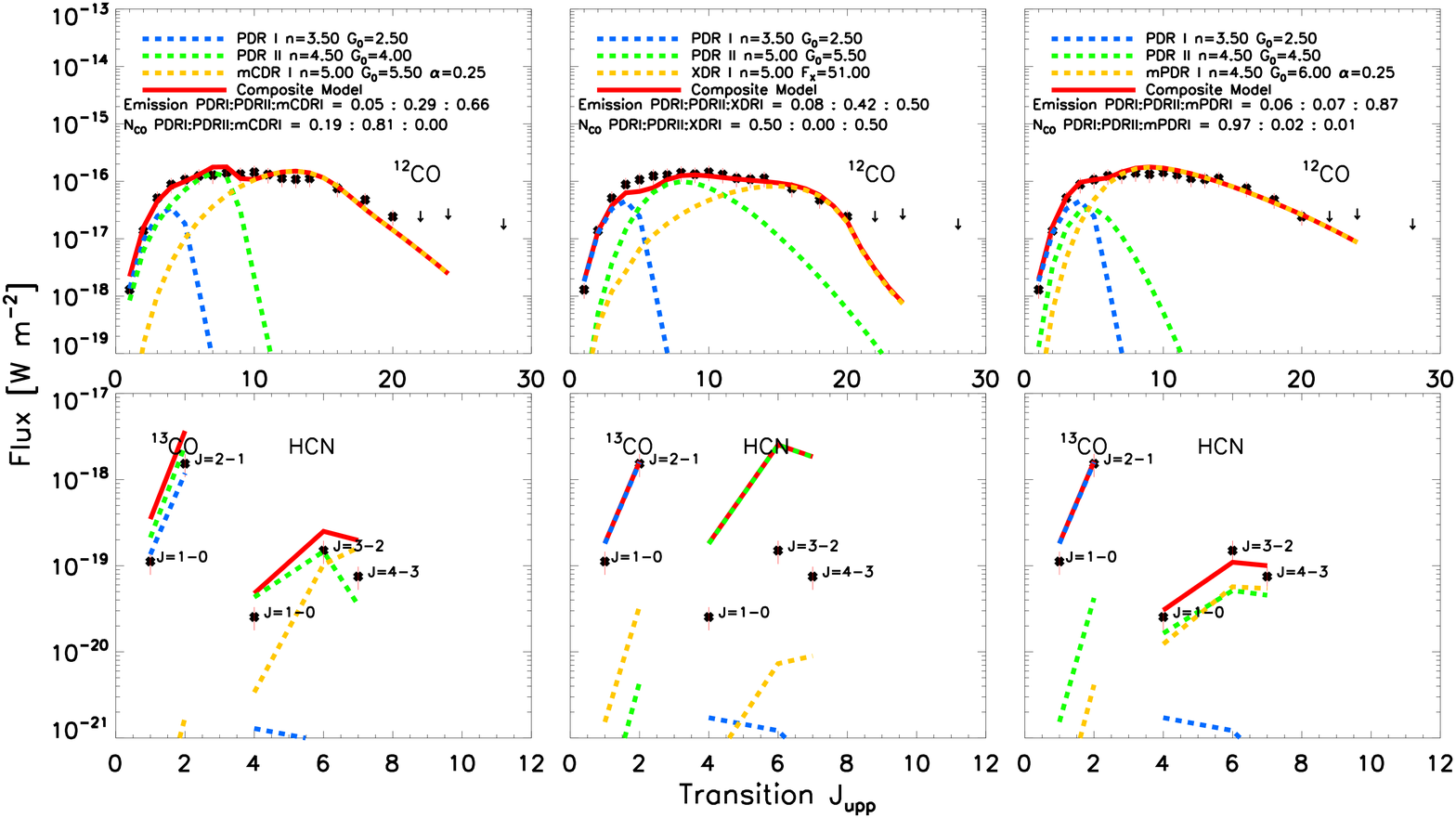}
\caption{$^{12}$CO (top), $^{13}$CO, and HCN (bottom) excitation ladders of Arp 299 A with the flux of each transition plotted as black asterisks with red error bars.  
In blue, green and yellow dotted lines we plot the PDR/PDR/mCDR (left), PDR/PDR/mPDR (center), and PDR/PDR/XDR(right) ISM phases with their filling factors.  
The composite model is plotted with a red solid line.  The model H$_2$ density [log cm$^{-3}$], $G/$G$_0$ [log Habing flux] or F$_X$ [log erg s$^{-1}$ cm$^{-2}$], and
percentage mechanical heating ($\alpha$) are shown in the legend along with the relative contribution of 
each phase in terms of emission and column density.  For emission, we compared the integrated intensity of each ISM phase to the total
modeled CO flux (the red line).  For the column density we perform the same calculation except comparing the column density of each phase to
the total combined model column density.}
\label{fig:all}
\end{figure*}

At the surface the heating budget is dominated by
photoelectric heating.  Both the mPDR and CDR 
models have the same basic radiative transfer and chemistry as
the PDR models, with either an enhanced cosmic ray rate or mechanical heating.  In the classical PDR models, the far-UV photons often do not penetrate far enough to affect the molecular region.  Thus, far-UV heating, cosmic ray heating, and mechanical heating can be varied in such a way that one source might dominate over the other depending on the depth into the cloud.  In the case of an enhanced cosmic ray ionization rate (CDRs), we increase the heating rate of the cosmic rays by a factor of 750 compared to the galactic value, used in the classical PDR models.  In the case of an added mechanical heating rate (mPDR), we add a new heating term to the heating balance of the classical PDR model, which we vary from 0-100\% of the UV heating at the surface of the PDR.  We use the names CDR and mPDR for convenience, to refer to PDR models with specific enhanced heating terms, yet both have the same classical PDR model base. 
On the other hand, X-rays heat gas in regions called X-ray dominated 
regions (XDRs), where the chemistry 
is driven by X-ray photons
instead of FUV photons \citep{2005A&A...436..397M}; the X-ray photons are able to penetrate farther into the cloud without efficiently heating 
the dust at the same time.  These X-rays are mostly produced by
active galactic nuclei (AGN) or in areas of extreme massive star formation and the strength of the X-ray radiation field (F$_{X}$)
is measured in erg s$^{-1}$ cm$^{-2}$. To test which excitation mechanisms are mainly responsible for heating the 
gas we will fit three cases.
\begin{enumerate}
 \item two PDRs one (m)CDR
 \item two PDRs one XDR
 \item two PDRs one mPDR
\end{enumerate}
We hold the first PDR and allow only the second and third ISM phases to vary, since PDR I is well constrained
using $^{13}$CO (Figure~\ref{fig:13}).  In case 1, we use the term (m)CDR since we allow the molecular emission
to be fit with either a pure CDR or a CDR with mechanical heating (mCDR). The best fit models are displayed for all three cases in Figure~\ref{fig:all}.  We give all model parameters for each case in
Table~\ref{tab:mcdr} and we discuss each case in detail below.

\begin{table*}
\caption{Model parameters for the three ISM phases for each of the three cases.}
 \begin{tabular}{|l||c|c|c|c|c|c|c|c|c|}
\hline
Component &  Density log(n$_{H}$) & log($G$) & log(N$_{CO}$) &log(N$_{H_2}$)& $\alpha$ & $\Omega$\tablefootmark{a} & C$_{em}$\tablefootmark{b} & C$_{N_{CO}}$\tablefootmark{c} & Mass$_{N_{H_2}}$\tablefootmark{d} \\
          &  log[cm$^{-3}$]  &  G$_0$  & log[cm$^{-2}$] &log[cm$^{-2}$]& \%& && &M$_\odot$ \\
\hline
\multicolumn{9}{|c|}{\textbf{Case 1}      M$_{tot}$: $3\times10^9$ M$_\odot$ } \\
\hline
PDR I   & 3.5 & 2.5 & 17.1 & 21.5&0 &0.9 & 0.05 & 0.19 & $2\times10^9$\\
PDR II  & 4.5 & 4.0 & 18.2 &21.9 &0 &0.3 & 0.29 & 0.81 & $1\times10^9$\\
mCDR I & 5.0 & 5.5 & 17.2  &21.0 &25 &0.006 & 0.66 &$<0.01$ &$3\times10^6$\\
\hline
\multicolumn{9}{|c|}{\textbf{Case 2}      M$_{tot}$: $4\times10^{9}$ M$_\odot$} \\
\hline
PDR I   & 3.5 & 2.5 & 17.1 & 21.5&0 &1.2 &0.11 & 0.50 & $2\times10^9$\\
PDR II  & 5.0 & 5.5 & 14.9 & 21.2&0&1.2 &0.40 & 0.003& $1\times10^9$\\
XDR I & 5.0 & 51.0\tablefootmark{e}& 19.4 &23.4 &0&0.006&0.48 & 0.50 &$1\times10^9$\\
\hline
\multicolumn{9}{|c|}{\textbf{Case 3}      M$_{tot}$: $3\times10^9$ M$_\odot$} \\
\hline
PDR I  & 3.5 & 2.5 & 17.1 & 21.5&0  &1.2& 0.06 & 0.97 & $2\times10^9$\\
PDR II & 4.5 & 4.5 & 15.5 &21.4 &0  &0.8 & 0.07& 0.02 & $1\times10^9$\\
mPDR I & 4.5 & 6.0 & 15.7 & 19.5&25&0.3 & 0.87 & $<0.01$ &$6\times10^6$\\
\hline
 \end{tabular}
  \tablefoot{ \\
 \tablefoottext{a}{$\Omega$ is the beam filling factor for each ISM phase.} \\
 \tablefoottext{b}{C$_{em}$ is the fractional contribution of each ISM phase to the emission (Eq.~\ref{eq:cem}).} \\
 \tablefoottext{c}{C$_{N_{CO}}$ is the fractional contribution of each ISM phase to the column density.} \\
 \tablefoottext{d}{Mass$_{N_{H_2}}$ is the mass of each ISM phase as estimated by the column density using Eq.~\ref{eq:col}.} \\
 \tablefoottext{e}{Units of XDR radiation field (F$_X$) are [erg s$^{-1}$ cm$^{-2}$].} \\
 }
\label{tab:mcdr}
\end{table*}

\subsubsection{Case 1}
In this case, the best fit is two PDRs and one mechanically heated CDR (mCDR) with 25\% mechanical heating at the surface.  We also tried a fit with one, two, and three pure CDRs.  The case of one CDR is the best of 
those options, yet has the same model parameters as the three PDR fits.  This suggests that it is still the UV photons that are heating the gas instead of a strong 
contribution from the 
cosmic rays.  In addition, in the case of one pure CDR, the HCN is overproduced by an order of magnitude, as in the three PDR fit.  Thus, we concentrate on the overall best
fit using an enhanced cosmic ray ionization rate and with 25\% mechanical heating.  This model is able to fit all $^{12}$CO transitions
within the error bars. However, the second and third ISM phases (green and yellow) produce more HCN luminosity than we observe.  The $^{13}$CO is also poorly fit, overproducing not only the J=1-0 but also the J=2-1 transition, thus making this an overall poor fit.  We suggest that cosmic rays 
play an unimportant role in heating the molecular gas in Arp 299 A, especially since the model needs 25\% mechanical heating in order to fit the high-J CO transitions.
  
\subsubsection{Case 2}
The second case that includes X-ray heating is justified by the existence of an AGN in Arp 299 A that could be heating a molecular torus around the AGN.  However, HCN is 
very poorly fit.  In order for an XDR to produce the high-J CO lines, it does not produce much emission in the mid-J CO lines.  This 
means that the mid-J CO lines must be produced by a powerful and dense PDR that results in bright HCN emission. Thus, the best fitting XDR model is the one that can reproduce
most of the high-J CO lines while producing minimal HCN emission. This points to the fact that XDR chemistry is unlikely to be the cause of the observed high-J CO emission. 
If it were responsible for the high-J CO emission, the third ISM phase would be the most massive one.  The mass of the XDR phase exceeds the total measured molecular mass
for the entire system, and thus we rule out X-rays as a significant heating source of the gas.  

\subsubsection{Case 3}
Case 3 represents two PDRs and one mechanically heated PDR. This case is the only case that fits all observed transitions within the error bars.  We also attempted to 
fit the observed transitions with one, two, and three mPDRs, yet in the case of two mPDRs, one of them has negligible mechanical heating, and in the case of three mPDRs there
is no reasonable fit. Thus, the situation represents a galaxy in which most of the gas is heated by UV photons, but a small amount of gas is heated almost entirely by 
mechanical heating. This gas could be in pockets of violent star formation where the stellar winds, jets, and/or supernovae are creating turbulence that efficiently couples to the gas.
The heating rate of the mechanical heating is $7.9\times10^{-19}$ erg s$^{-1}$ cm$^{-2}$, which represents $\sim 4\%$ of the total heating, reflecting the fact 
that in this system the mechanical heating is very localized.  We conclude that mechanical heating is the most likely candidate
for the additional heating source in Arp 299 A. 

This result is similar to what was found in NGC 253 \citep{2014arXiv1401.4924R,2008ApJ...689L.109H}, where mechanical heating is needed as an additional heating mechanism.  However,
in NGC 253, the system requires mechanical heating in all three ISM phases to reproduce the extremely bright CO emission.  Since Arp 299 A only needs mechanical heating to explain
the third, most extreme, ISM phase, this lends itself to isolated and localized mechanical heating deriving from either supernova remnants or extreme star formation regions with powerful winds.  The 
nuclear region of NGC 253
on the other hand, has universally bright CO lines, meaning the mechanical heating must be distributed throughout the galactic nucleus, perhaps coming from the massive molecular 
outflow \citep{2013Natur.499..450B, 1985ApJ...299L..77T}.  In addition, although the far-infrared luminosity in Arp 299 is about an order of magnitude higher than in NGC 253, we see much 
brighter cooling lines in NGC 253.  This is most likely due to a distance effect.  With SPIRE's beam size, in NGC 253 we observe only the nuclear region, while in Arp 299 A we 
observe the nucleus and surrounding disk.  Thus the extreme environment of the galactic nucleus is averaged out with the less luminous disk regions in Arp 299.  

\subsection{Molecular gas mass}

We can estimate the mass of each molecular gas ISM phase as well as the total molecular gas mass for each case using Equation \ref{eq:col}.  We find a total
molecular mass equal to $3\times10^9$ M$_\odot$ regardless of excitation mechanism, which is in good agreement with the literature values.  In addition, using the dust
mass of 1.1$\times10^8$ M$_\odot$, we find a gas to dust ratio of $\sim 30$. The first ISM phase is well constrained in all cases; the mass
is $2\times10^9$ M$_\odot$, making it the most massive component in most cases. The second ISM phase is also well constrained with a 
mass of $1\times10^9$ M$_\odot$.  The third ISM phases' mass is not as well constrained, yet it is the least massive component in most cases, ranging
from $3-6\times10^6$ M$_\odot$, except in the case of the XDR, where this phase is more massive, $1\times10^9$ M$_\odot$.  

The fact that the mass is so well constrained underlines the importance of observing even just two $^{13}$CO transitions.  We also see the strength of including
HCN measurements to constrain both the high density ISM phase, and the excitation mechanism.  Further, these results agree with those from \citet{2014arXiv1401.4924R}
that mechanical heating plays an important role in understanding the molecular line emission, even though UV heating is still the most dominant heating source.

\section{Limitations and usefulness of the $^{12}$CO ladder}
\label{sec:limit}
Herschel SPIRE gave access to the full CO
ladder ranging from J=4-3 to J=13-12, in the nearby universe.  With
Herschel PACS, higher J lines could also be observed.  Before
Herschel, it was thought that observing the flux of CO
transitions greater than J=10 would break the degeneracy between UV excitation
and X-ray excitation.  However, now that
the wealth of observations from
the Herschel Space Observatory are available, access to the full CO ladder does not
necessarily break this degeneracy. In fact, the information that can be
extracted from observations of only $^{12}$CO is very limited.

Qualitatively, bright $^{12}$CO emission indicates the presence of warm
molecular gas.  However, without any other information, the source of
heating, the amount (mass) of heated gas, and the precise density and temperature cannot be
determined.  It is possible, however, to extract the
turnover point of the $^{12}$CO.  If the turnover point is in the low to mid-J
transitions (from J=1-0 to J=6-5), as seen in Arp 299 B+C, the gas is most likely heated by
UV photons in PDRs.  If the turnover is higher than that, it can be
either an extreme PDR ($n_H>10^5$ cm$^{-3}$, $G>10^5$ G$_0$), X-rays, cosmic rays, or mechanical heating that
may be responsible.  This is demonstrated fitting a pure PDR model to the high-J CO lines, and is clear in the 
degeneracy parameter space diagrams shown in Figure 3 by \citet{2014arXiv1401.4924R}.   

For metal-rich extragalactic sources, the $^{12}$CO ladder
represents all the molecular clouds in the galaxy, spanning a range of
physical environments.  Therefore, multiple ISM phases are necessary to
fit the ladder.  These ISM phases represent many clouds with similar
physical properties.  In addition, with just transitions of $^{12}$CO,
you can determine distinct density-temperature combinations
for each phase. In general, the low-J CO lines are from a lower density,
lower temperature ISM phase, but the density-temperature combination is highly degenerate.  The mid-J CO transitions arise from a
warm and medium density phase, and the high-J transitions from a high
density, high temperature phase.

If multiple transitions of $^{13}$CO are added, then the beam averaged optical
depth, and thus column density, are constrained.  This allows for a
better constrained mass estimate.  It also helps lessen the
temperature-density degeneracy, but does not break it.  In order to break this degeneracy, other molecules must be added.  For
example, HCN, HNC, and HCO$^+$ are good tracers of density for high
density environments.  For lower density regimes, [CI] can be a good probe of
the gas temperature, yet it is very difficult to interpret, since we cannot disentangle 
different emitting regions within our beam. To
summarize, if bright $^{12}$CO emission is observed, there is
warm gas.  Yet in order to probe the physical parameters of that gas,
other molecular information is crucial. Some of these molecules do not always originate from the
same spatial location as the $^{12}$CO and may be tracing a different gas
component altogether. Interpreting $^{12}$CO is not trivial
and the analysis should be performed with an understanding of the
challenges and limitations.

Many Herschel SPIRE CO ladders have been obtained from luminous infrared galaxies, and they all require 
some additional heating mechanism to explain the high-J CO emission.  For example in Arp 220, \citet{2011ApJ...743...94R} find that PDRs, XDRs, and CDRs can be ruled out, while 
the mechanical energy available in this galaxy is sufficient to heat the gas. Similarly, \citet{2013ApJ...762L..16M} find strong evidence for
shock heating in NGC 6240.  On the other hand, \citet{2012ApJ...758..108S} and \citet{2010A&A...518L..42V} find in NGC 1068 and Mrk 231 respectively, that it is likely 
XDR heating responsible for the high excitation CO lines.  Although both of these sources have confirmed AGN, the CO ladder fitting was not combined with 
a dense gas tracer (HCN/HNC/HCO$^+$), and thus mechanical heating cannot be directly ruled out. The picture emerging from the SPIRE CO-ladders is 
that in these extreme star forming galaxies, the gas is rarely heated by only UV photons and that in most cases, the molecular gas is heated through 
either X-rays or mechanical heating. 

\section{Conclusions}
\label{sec:conc}
We observed Arp 299 with Herschel PACS and SPIRE in both the spectrometer and photometer mode.  The Herschel SPIRE FTS observations had three 
separate pointings, namely towards
Arp 299 A, B and C.  The pointings of Arp 299 B and C are overlapping so it is
difficult to separate the emission from each nucleus. We extract the
line fluxes of the CO transitions, [CI],
[NII], and bright H$_2$O lines for Arp 299 A, B, and C separately. We also measure the continuum fluxes
from SPIRE photometer mode at 250, 350, and 500 $\mu$m.  With PACS, we detect CO transitions 
from J=14-13 to 20-19 and upper limits up to J=28-27.  Using these data, we find:   
\begin{enumerate}
\item A simple quantitative comparison of 
the spectra of Arp 299 A with B and C shows that the environment of source A is 
much more excited, with more warm molecular gas.  

\item Using the full range of CO transitions we construct CO excitation ladders
for each of the three pointings.  Again, the CO ladders reveal a clear
disparity between Arp 299 A and B+C; source A displays a flattened ladder, while B+C turns over around J=5-4.  

\item Since we have high-J $^{12}$CO PACS observations along with $^{13}$CO and HCN JCMT observations of Arp 299 A, we perform 
an automated $\chi^2$ minimized fitting routine to fit the CO and HCN ladders with three ISM components.  We find a suitable 
fit for $^{12}$CO and $^{13}$CO but not for HCN. In addition, the third
ISM phase would then be a truly extreme PDR, an order of magnitude more extreme than Orion Bar. 

\item We create an infrared SED using values from the literature along with PACS and SPIRE continuum measurements.  We fit this SED with
three gray bodies and determine the temperature and mass of each dust component (cold, warm, and hot).  We do not observe enough hot dust to match 
the amount of hot dust that would then be produced by the extreme PDR, in the case of a fit by three pure PDRs. Thus, we conclude that the flattening of the CO ladder, 
and extra excitation of the $^{12}$CO in Arp 299 A in comparison to B+C, is due 
to an additional heating mechanism. 

\item We allow the third ISM phase (high density, high excitation) to have additional heating by cosmic rays, mechanical heating, and X-rays.  We find 
mechanical heating to be the most likely additional heating source since it fits all transitions within the errors.  As the best fit model requires 
mechanical heating only in the third component, this suggests that for Arp 299A the mechanical heating is localized, likely to come from supernovae remnants or
pockets of intense star formation. 

\item We caution the use of $^{12}$CO alone as a tracer of the physical conditions of the ISM. We find that $^{12}$CO reveals only the presence of 
warm molecular gas, but that the amount, physical properties, and heating source cannot be determined without observations of other molecules. 

\end{enumerate}

\begin{acknowledgements}
The authors would like to thank the referee for their fruitful feedback and time
spent. The authors gratefully acknowledge financial support under the ``DeMoGas'' project. 
The project ``DeMoGas'' is implemented under the ``ARISTEIA'' Action of the ``PERATIONAL PROGRAMME EDUCATION 
AND LIFELONG LEARNING'' and is co-funded by the European Social Fund (ESF) and National Resources.
We would like to thank Edward Polehampton for his help preparing the SPIRE observations.  SPIRE has been developed by a consortium of institutes led by
Cardiff Univ. (UK) and including: Univ. Lethbridge (Canada);
NAOC (China); CEA, LAM (France); IFSI, Univ. Padua (Italy);
IAC (Spain); Stockholm Observatory (Sweden); Imperial College London, RAL, UCL-MSSL, UKATC, Univ. Sussex (UK);
and Caltech, JPL, NHSC, Univ. Colorado (USA). This development has been supported by national funding agencies:
CSA (Canada); NAOC (China); CEA, CNES, CNRS (France);
ASI (Italy); MCINN (Spain); SNSB (Sweden); STFC, UKSA
(UK); and NASA (USA). The Herschel spacecraft was designed, built, tested, and 
launched under a contract to ESA managed by the Herschel/Planck Project team by 
an industrial consortium under the overall responsibility of the prime contractor 
Thales Alenia Space (Cannes), and including Astrium (Friedrichshafen) responsible 
for the payload module and for system testing at spacecraft level, Thales Alenia 
Space (Turin) responsible for the service module, and Astrium (Toulouse) responsible 
for the telescope, with in excess of a hundred subcontractors. HCSS / HSpot / HIPE is 
a joint development (are joint developments) by the Herschel Science Ground 
Segment Consortium, consisting of ESA, the NASA Herschel Science Center, and 
the HIFI, PACS and SPIRE consortia.

\end{acknowledgements}

\bibliographystyle{aa}
\bibliography{bib_file.bib}

\end{document}